\documentclass{aa}  
\usepackage{graphicx}
\usepackage{color}
\usepackage{hyperref}
\usepackage{float}
\usepackage{txfonts}
\usepackage{xcolor}
\usepackage{natbib}
\usepackage{epsf} 

\def \ME{M$_{\oplus}$}

\def \MS{M$\mathrm{_\odot}$}

\def \1s{$1\,\sigma$}

\def \t0{T$_0$}

\def \wapiti{\texttt{wapiti}}
\def \radvel{\texttt{RadVel}}

\begin{document}

\title{Characterizing planetary systems with SPIRou: a temperate sub-Neptune exoplanet orbiting the nearby fully-convective star GJ~1289 and a candidate around GJ~3378\thanks{Based on observations obtained at the Canada-France-Hawaii Telescope (CFHT) which is operated by the National Research Council (NRC) of Canada, the Institut National des Sciences de l'Univers of the Centre National de la Recherche Scientifique (CNRS) of France, and the University of Hawaii. The observations at the CFHT were performed with care and respect from the summit of Maunakea which is a significant cultural and historic site. } }

    \author{
C. Moutou\inst{1} \and
M. Ould-Elhkim\inst{1} \and 
J.-F. Donati\inst{1} \and
P. Charpentier\inst{1} \and 
C. Cadieux\inst{2} \and 
X. Delfosse\inst{3} \and
E. Artigau\inst{2} \and
L. Arnold\inst{4}  \and 
C. Baruteau\inst{1} \and
A. Carmona\inst{3} \and 
N.J. Cook\inst{2} \and
P. Cortes Zuleta\inst{5} \and
R. Doyon\inst{2} \and
G. Hébrard\inst{6} \and the SLS consortium
}

\institute{
\inst{1} Univ. de Toulouse, CNRS, IRAP, 14 avenue Belin, 31400 Toulouse, France, \email{claire.moutou@irap.omp.eu} \\
\inst{2} Institut Trottier de Recherche sur les Exoplanètes, Université de Montréal, 1375 Ave Thérèse-Lavoie-Roux, Montréal, QC H2V 0B3, Canada\\
\inst{3} Univ. Grenoble Alpes, CNRS, IPAG, 38000 Grenoble, France\\
\inst{4} Canada-France-Hawaii Telescope, CNRS, 96743 Kamuela, Hawaii, USA\\
\inst{5} Aix-Marseille Université, CNRS, CNES, LAM (Laboratoire d’Astrophysique de Marseille), Marseille, France \\
\inst{6} Institut d'Astrophysique de Paris, UMR7095 CNRS, Universit\'e Pierre \& Marie Curie, 98 bis boulevard Arago, 75014 Paris, France \\
}

\date{}

\abstract{We report the discovery of two new exoplanet systems around fully convective stars, found from the radial-velocity (RV) variations of their host stars measured with the nIR spectropolarimeter CFHT/SPIRou over multiple years. GJ~3378 b is a planet with minimum mass of $5.26^{+0.94}_{-0.97}$ \ME\ in an eccentric 24.73-day orbit around an M4V star of 0.26 \MS. GJ~1289 b has a minimum mass of $6.27\pm1.25$ \ME\ in a 111.74-day orbit, in a circular orbit around an M4.5V star of mass 0.21 \MS. Both stars are in the solar neighbourhood, at respectively 7.73 and 8.86 pc. The low-amplitude RV signals are detected after line-by-line post-processing treatment. These potential sub-Neptune class planets around cool stars may have temperate atmospheres and be interesting nearby systems for further studies. We also recovered the large-scale magnetic field of both stars, found to be mostly axisymmetric and dipolar, and with a polar strength of 20-30~G and 200-240~G for GJ 3378 (in 2019-21) and GJ 1289 (in 2022-23), respectively. The rotation periods measured with the magnetic field differ from the orbital periods, and in general, stellar activity is not seen in the studied nIR RV time series of both stars. GJ~3378~b detection is not confirmed by optical RVs and is therefore considered a candidate at this point. }
\keywords{ Planetary systems -- Techniques: radial velocities -- Instrumentation: spectrographs -- Individual stars: GJ~1289, GJ~3378}

\titlerunning{The SPIRou planet-search program }
\authorrunning{Moutou et al}
\maketitle

\section{Introduction}
\label{sec:1}
Although M dwarfs are the most common stars in the Galaxy \citep{reyle2021}, the exoplanet population accompanying these low-luminosity parent stars can only be closely studied within the solar neighbourhood. In the last decade, it appeared that, apart from a few exemplary systems which contain a massive planet, the trend for low-mass stars is to have low-mass planetary companions \citep{bonfils2013, dressing2015, sabotta2021}. This configuration is also confirmed by exoplanet population synthesis, as recently developed in \citet{burn2021}. 

The population of planets around the latest type stars is, however, less constrained with observations than the earlier types. Such stars are faint, sometimes rapidly rotating and active, which complicates their observation. However, a few famous planetary systems in this category have been discovered in the last decade, like those of Proxima Cen \citep{anglada-escude2013}, Trappist-1 \citep{gillon2016}, or GJ~1002 \citep{suarez2023}.

Precision radial-velocity (RV) measurements of M dwarfs were first obtained in the optical domain then, more recently, the spectral range of such observations has been extended towards the far-red and near-infrared (nIR) domains. A first advantage is to have instruments more sensitive to cool, red stars \citep{artigau2018b}. In addition, getting RV observations in a wide spectral range helps distinguishing between chromatic stellar effects and achromatic planetary signatures \citep{huelamo2008, reiners2013, carmona2023}. When optical and nIR spectra are obtained at the same time, it also expands the number and diversity of observed spectral lines, as with HARPS+NIRPS \citep{wildi2022}, CARMENES \citep{quirrenbach2014}, or GIARPS \citep{claudi2016}. 

RV measurements in the nIR domain, however, may be plagued with systematic noise at a higher level than optical RVs, mostly due to the presence of telluric lines and the varying relative position of the telluric and stellar spectra in the instrumental focal plane. More stringent corrections of the extracted spectra are needed, compared to optical data, to get rid of this telluric contamination. Then, specific corrections at the level of stellar lines are required to remove any residual signature of telluric signals affecting the RV time series. Advanced techniques have been developed in the last years based on line-by-line analyses \citep{cretignier2021, artigau2022, ouldelhkim2023}.

Low-mass stars tend to be more active due to the properties of their magnetic field \citep{morin2010,kochukhov2021,mignon2023}. Recent studies even suggest that some of them are able to trigger unusually strong large-scale magnetic fields despite their long rotation periods \citep{lehmann2024} and that their flaring rate remains high \citep{medina2022}. While requiring caution for the interpretation of RV signals whose period is close to the rotation period (and first harmonics), this does not prevent us from searching for planetary companions around such stars.

In this paper, we investigate the RV time series of two fully-convective stars observed in the nIR domain with SPIRou \citep{donati2020}.
Two new exoplanets are found around the nearby stars GJ~3378 and GJ~1289 after an intensive RV campaign. In section 2, the target stars and the SPIRou observations are described as well as the data analysis method and the magnetic properties of the parent stars. In section 3, the new detected RV signals are presented. In section 4, we discuss these new planets in the context of M-star exoplanet populations and conclude.

\section{Observations}
\label{sec:2}
\subsection{SPIRou data}
High-resolution spectroscopy and RV measurements were performed at the Canada-France-Hawaii observatory with the SPIRou (SpectroPolarimetre InfraROUge) instrument \citep{donati2020}. SPIRou covers the 985-2450 nm wavelength range with a 70000 resolving power and is highly stabilised for precise RV measurements. Data were collected in the framework of the SPIRou Legacy Survey Planet Search program described in \citet{moutou2023}\footnote{Program ID P40 in semesters 19A to 22A}. For GJ~1289, additional data were collected within the SPICE large program\footnote{Program ID P45 from semesters 22B to 24A} in continuation, with homogeneous quality.

SPIRou performs velocimetric and spectropolarimetric measurements simultaneously. Each visit consists of four consecutive exposures with different positions of the quarter-wave plates, in order to achieve a circular polarisation (Stokes $V$) analysis of the stellar light. Spectropolarimetric measurements of the target stars (up to BJD 2459800) were described and analysed in previous work \citep{fouque23,donati2023,lehmann2024}. The RV measurement, in turn, is done on individual exposures first (details below), and then combined in a single per-night RV value.

GJ~3378 was observed with SPIRou 181 times between 18 September 2019 and 31 January 2022, over a time span of 876 days. Airmass ranges from 1.301 to 2.037. The spectrum moves within a Barycentric Earth RV (BERV) range of $\pm$ 24.5 km/s due to the Earth motion around the Sun, a configuration that is favourable to getting a clean stellar template and good performance in telluric correction (Ould-Elhkim et al, in prep.). The average exposure time is 544s per visit and the mean obtained S/N is 212 per pixel in the $H$ band.\\
GJ~1289 was observed with SPIRou 300 times between 15 May 2019 and 25 January 2024 with a time span of 1686 days. Airmass ranges from 1.05 to 2.54 and observations have a maximum BERV range of $\pm$ 25.2 km/s. Average exposure times and S/N per nightly visit are, respectively, 800s and 200 per pixel in the $H$ band. Tables of observational data and derived values are given online.

\subsection{Stars}
GJ~3378 and GJ~1289 are both fully-convective M stars located at 8.86 and 7.73~pc, respectively \citep{GaiaDR3} -full-convective stars being defined by a mass lower than 0.35\MS\ \citep{chabrier1997}. Their metalicity, close to the solar one, and their estimated mass, obtained from the spectroscopic analysis of SPIRou spectra by \citet{cristofari2022} and \citet{cristofari2023}, are summarised in Table \ref{tab:1}. The rotation periods have been measured using the quasi-periodic modelling of their longitudinal magnetic-field modulation. GJ~1289 has a rotation period of 73.66$\pm$0.92 days while GJ~3378 has a rotation period of 95.1$\pm$2.3 days \citep{donati2023}. Note that \citet{sabotta2021} reports a shorter rotation period for GJ~3378, of 83.4 days from the variation of the average stellar line width, possibly suggesting differential rotation at the surface of the star.
Despite a slow rotation, GJ~1289 hosts a strong large-scale magnetic field that rapidly evolves with time, with, for instance, a dipole strength increasing from 80 to 200 G from 2019 to 2021 \citep{lehmann2024}. More recent data included in this paper are described and analysed in section \ref{topo}.
The magnetic field of GJ 3378 is clearly detected as shown in \citet{donati2023} and its topology is described in section \ref{topo}.

The small-scale magnetic fields of GJ 3378 and GJ 1289 were also estimated from their mean intensity spectra, which were calculated from the same collection of SPIRou spectra by \citet{cristofari2023} using \texttt{Zeeturbo}. A mean field of 150$\pm$50 G was found at the surface of GJ 3378, picturing a very quiet star with 94\% of its surface at undetected levels of magnetic field, and active regions totalling only 6\% of the surface at 2kG. GJ 1289 has a stronger field, with a mean value of 1010$\pm$80 G and 45\% of its surface at more than 2kG. The time dependence of the small-scale magnetic field of GJ 1289 will be studied in a forthcoming paper (Charpentier et al, in prep).

\begin{table}
    \centering
\caption{Stellar parameters of GJ~3378 and GJ~1289}
\label{tab:1}
\begin{tabular}{lrrl}
Star        & GJ 3378 & GJ 1289 & Ref.\\\hline
distance (pc) & 7.73 & 8.86 & 1\\
Mass (\MS)  & 0.26$\pm$0.02 & 0.21$\pm$0.02 & 2\\
Sp Type     & M4V & M4.5Ve & 1\\
M/H        & -0.05$\pm$0.10& 0.02$\pm$0.10 & 3\\
T$_{eff}$ (K)  & 3378$\pm$30 & 3296$\pm$30 & 3\\
$<B>$ (G) & 150$\pm$50 & 1010$\pm$80 & 3\\
$P_{\rm rot}$ (d) & 95.1$\pm$2.3 days& 73.66$\pm$0.92 & 4 \\\hline
\end{tabular}
\tablebib{(1)~\citet{GaiaDR3}; (2) \citet{cristofari2022}; (3) \citet{cristofari2023}; (4) \citet{donati2023}.}
\end{table}

\subsection{Magnetic topology of parent stars}
\label{topo}

We first applied Least-Squares Deconvolution \citep[LSD;][]{donati1997} to all recorded Stokes $V$ spectra of GJ~3378 and GJ~1289, using a M3V mask as described in \citet{donati2023}.  We then analysed the time series of Stokes $V$ LSD profiles within each observing season, translating the observed rotational modulation into a map of the large-scale field using Zeeman-Doppler Imaging \citep[ZDI;][]{donati2006}.  We proceeded as outlined in \citep{lehmann2024}, assuming that the large-scale magnetic topology is constant over each observing season, setting the inclination angle of the stellar rotation axis to the line of sight to $i=60\degr$, and using the rotation period derived from the quasi periodic variations of the longitudinal field \citep[][see Table \ref{tab:1}]{donati2023}.  

\begin{figure*}[ht!]
    \centering
\includegraphics[width=0.9\hsize]{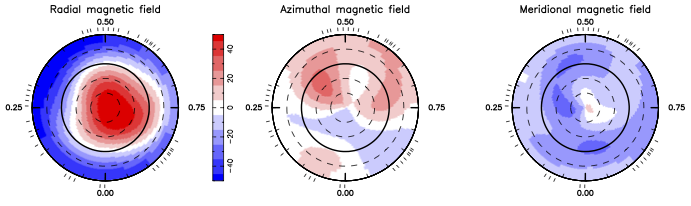}
\includegraphics[width=0.9\hsize]{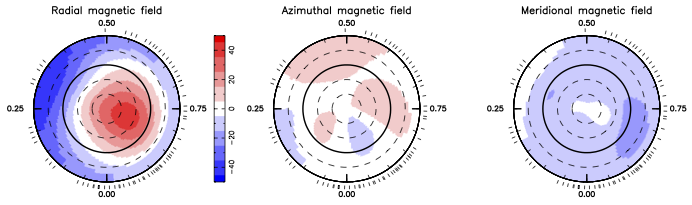}
\includegraphics[width=0.9\hsize]{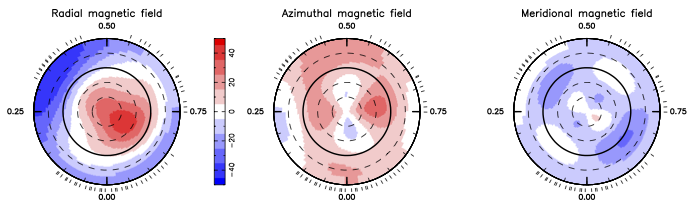}
\caption{Reconstructed maps of the large-scale field of GJ\,3378 in seasons 2019, 2020, 2021 (from top to bottom). The left, middle and right columns represent the radial, azimuthal, and meridional field components in spherical coordinates. The colour bar is expressed in Gauss. The maps are shown in a flattened polar projection down to latitude --60\degr\, with the  pole at the centre and the equator depicted as a bold line. Outer ticks indicate phases of observations, assuming a rotation period of 95.1~d and counting from an arbitrary BJD of 2458745.0.}\label{fig:map3378}
\end{figure*}

\begin{figure*}[ht!]
    \centering
\includegraphics[width=0.9\hsize]{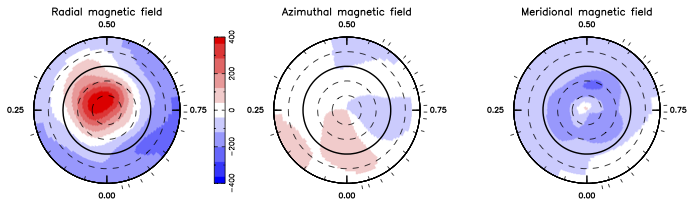}
\includegraphics[width=0.9\hsize]{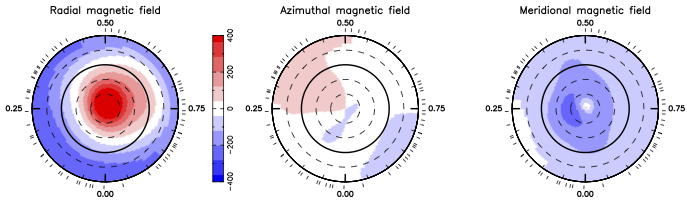}
\caption{Same as Fig.~\ref{fig:map3378} for GJ\,1289 in seasons 2022 (top) and 2023 (bottom). Phases are computed assuming a rotation period of 73.66~d and counting from an arbitrary BJD of 2458649.0.}\label{fig:map1289}
\end{figure*}

The large-scale magnetic maps we obtain are shown in Figs.~\ref{fig:map3378} and \ref{fig:map1289} for GJ~3378 (in seasons 2019/20, 2020/21 and 2021/22) and GJ~1289 (2022/23 and 2023/24) respectively, with the main characteristics of the recovered fields being listed in Tables~\ref{tab:tab3378} and \ref{tab:tab1289}. For all seasons and for both stars, the reconstructed topologies are mainly poloidal, as expected from the low sensitivity of Stokes $V$ profiles to toroidal fields in stars as slowly rotating as our targets \citep{lehmann2024}, whereas the poloidal fields are mostly dipolar (with the dipole enclosing 85--95\% of the magnetic energy). Moreover, the reconstructed fields are largely axisymmetric, with the magnetic dipoles being tilted at 20--38$\degr$ to the rotation axis.  We find that the large-scale field of GJ~3378 (in the range 20--30~G, see Table~\ref{tab:tab3378}) is weakening with time over our 3 observing seasons, whereas the much stronger field of GJ~1289 (in the range 200--250~G, see Table~\ref{tab:tab1289}) continues the strengthening trend already pointed out in \citet{lehmann2024}.  No major change in the overall polarity of the large-scale fields \citep[such as those reported for similar M dwarfs in][]{lehmann2024}, nor of their main characteristics, is observed for these two targets over our long-term observing campaign.  In Table~\ref{tab:tab3378}), we list for each observing season the average reconstructed large-scale field strength (<$B_V$>), the polar strength of the dipole component ($B_{\rm dip}$), the amount of magnetic energy reconstructed in the poloidal ($\mathrm{f}_{\rm pol}$) and axisymmetric ($\mathrm{f}_{\rm axi}$) components of the field, the tilt of the dipole field to the rotation axis, and the rotation phase towards which the dipole points (in the ephemeris mentioned on Fig.~\ref{fig:map3378}).  Error bars are typically equal to 5--10\% for field strengths and percentages, and 5--10\degr\ for field inclinations.

\begin{table}
\caption{Properties of the large-scale magnetic field of GJ~3378}
\begin{tabular}{lccc}
\hline
Season             & 2019/20 &2020/21& 2021/22\\\hline
<$B_V$> (G)           & 31     &  22     &   21   \\ 
$B_{\rm dip}$ (G)         & 47       &  33     &   30 \\
$\mathrm{f}_{\rm pol}$ (\%)     &  100   &  100    &  100  \\
$\mathrm{f}_{\rm axi}$ (\%)     &  88   &  63     &  70  \\
Dipole tilt angle ($\degr$)  & 20  & 38  &  30  \\
Pointing phase     &  0.79   &    0.80   &   0.82 \\
Nb. obs            &   40      &   72    &   64 \\\hline
\end{tabular}
\label{tab:tab3378}
\end{table}

\begin{table}
\caption{Properties of the large-scale magnetic field of GJ~1289}
\begin{tabular}{lcc}
\hline
Season             & 2022/23 &2023/24\\\hline
<$B_V$> (G)           &   204      &     243       \\ 
<$B_{\rm dip}$> (G)         &   276      &     344      \\
$\mathrm{f}_{\rm pol} (\%)  $   &   100      &     100      \\
$\mathrm{f}_{\rm axi} (\%)  $   &   78      &    85       \\
Dipole tilt angle ($\degr$)  & 30  &  26     \\
Pointing phase     &   0.37  &   0.63        \\
Nb. obs            &    18     &    69       \\\hline
\end{tabular}
\label{tab:tab1289}
\end{table}

\subsection{Radial velocity measurements and post-processing treatment}
The precision RV measurements are obtained from an extraction of the spectra with the SPIRou pipeline APERO \citep{cook2022} followed by the line-by-line (LBL) analysis developed for SPIRou by \citet{artigau2022}.\\
The PCA-based line selection method called \wapiti\footnote{https://github.com/HkmMerwan/wapiti} \citep{ouldelhkim2023} is then used on LBL time series to correct for telluric residuals as is relevant with nIR RVs and small signals. The correction is only applied to the lines that are more affected by residual noise. There are three steps in the post-treatment, as described by \citet{ouldelhkim2023}: 
\begin{itemize}
    \item The selection of all spectra with S/N larger than 70\% of the requested S/N and all lines present in at least 50\% of the epochs. The reason behind this first cut-off is that PCA is highly sensitive to outlier measurements strongly increasing variance and thus affecting the first components. Rejecting lines that are affected by tellurics most of the time is also an important cleaning step, and regularly done in precision RV measurements using masks.
    \item The removal of the time-dependent zero point RVs common to the 19 most RV-quiet stars of the SLS sample \citep{moutou2023}.
    \item The reconstruction and correction of the systematic noise using weighted PCA modelling
\end{itemize}

\begin{figure*}[h!]
    \centering
\includegraphics[width=0.49\hsize]{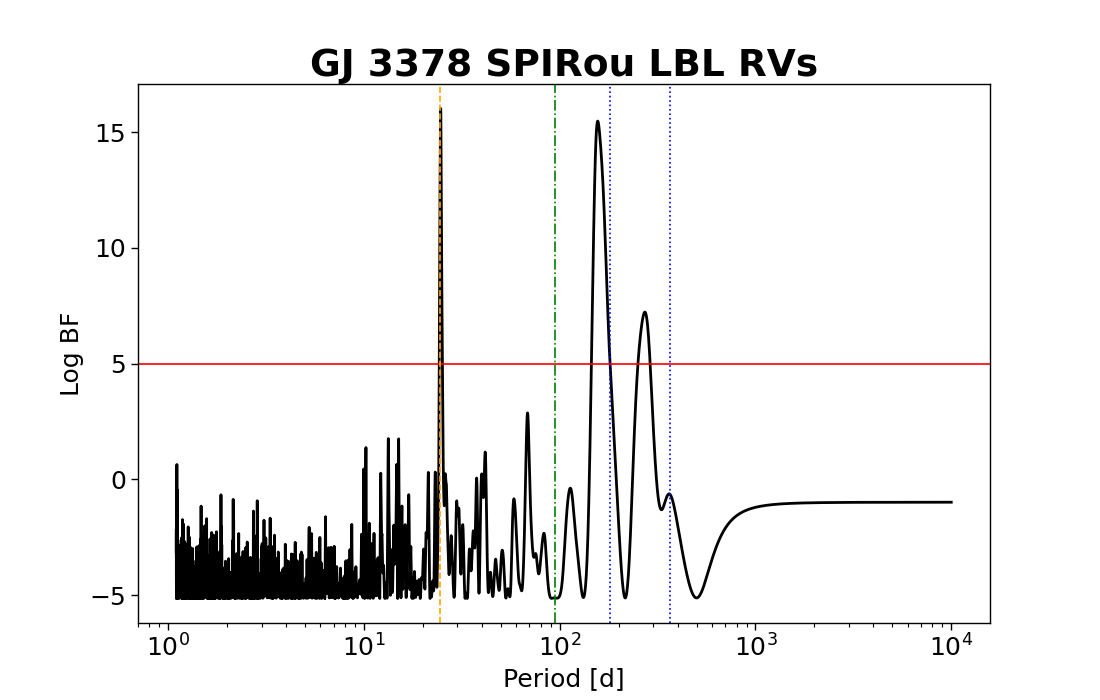}
\includegraphics[width=0.49\hsize]{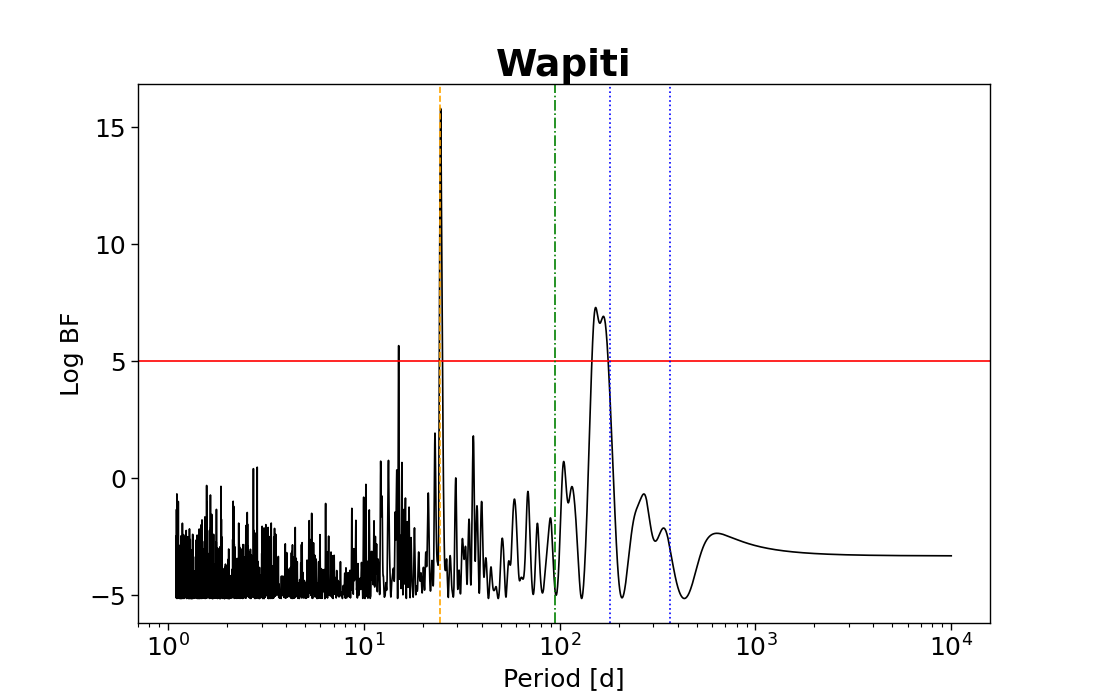}
\includegraphics[width=0.49\hsize]{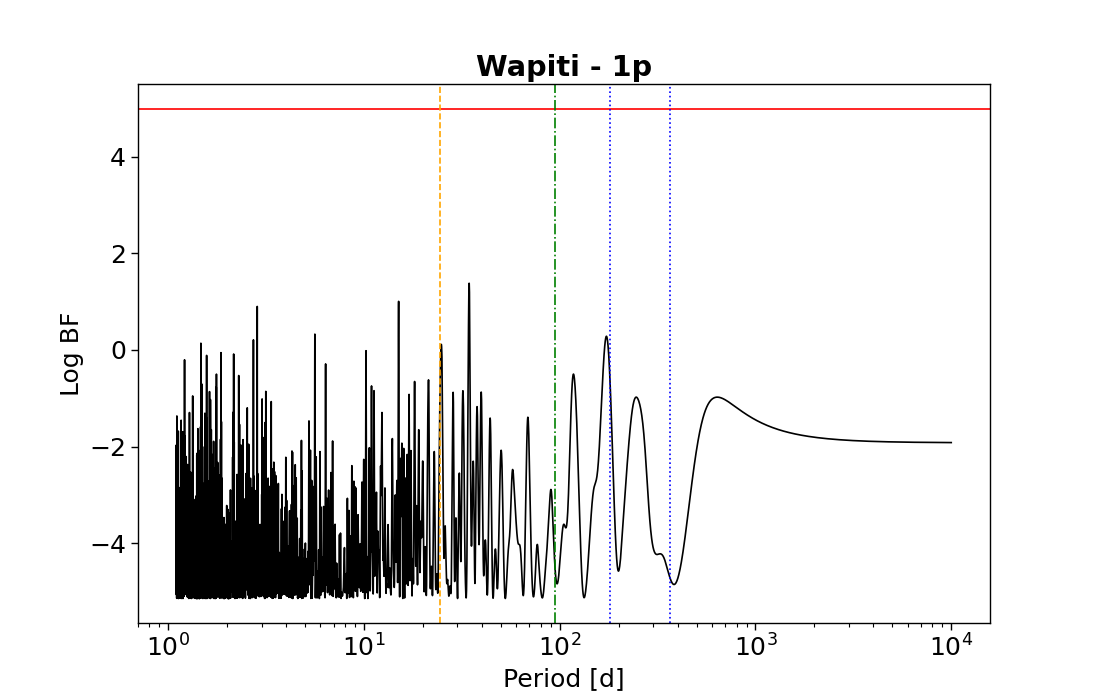}
\includegraphics[width=0.49\hsize]{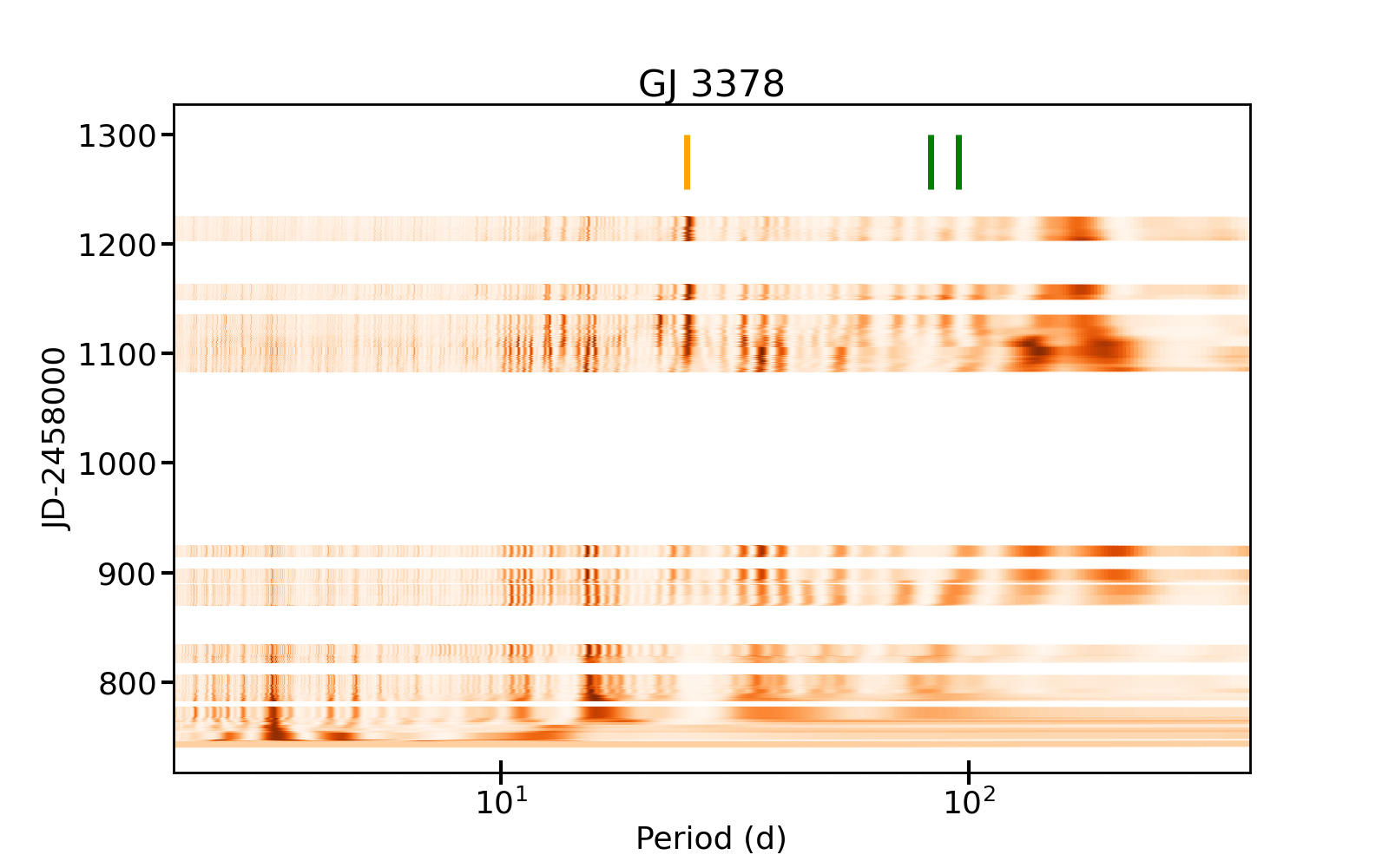}
\caption{Periodograms of GJ~3378. Top left: SPIRou RVs after pre-cleaning (drift, outlier, and most affected lines removed); top right: \wapiti-corrected RVs (wPCA corrected); bottom left: residuals after one planet model is removed; and bottom right: stacked periodogram. The blue dotted lines show 1-year and half-year periods. The green dot-dash line shows the rotation period as measured with the magnetic field proxy. The orange dashed line shows the orbital period, and the red horizontal line shows the detection threshold of log$Bf=5$. On the stacked periodogram, the Y axis corresponds to the time; the orange and green lines show the orbital (24.7d) and rotational (83-95d) periods, respectively.}\label{fig:1}
\end{figure*}

For GJ~3378 and GJ~1289, the number of ordered noisy components that were removed are, respectively, 4 and 12. These noisy components clearly show a correlated behaviour with the BERV, symptomatic of telluric residuals. However, the exact number of corrected components in these two cases has no critical impact on the result, since the signal detection is already significant without the \wapiti\ treatment (Figures \ref{fig:1} and \ref{fig:2}). This additional correction, however, allows reducing the one-year signal due to telluric residuals and its harmonics, and may boost any other signal not related to systematic noise. This has shown to be efficient in the case of known planet signals \citep{ouldelhkim2023}. The role of the zero-point correction could also be important (although it has been minor for SPIRou observations so far): other instrumental noises, like tiny spectral shifts of the wavelength solution at the m/s level, and common to all measurements may need to be corrected for, as is common practice for other spectrographs \citep{courcol2015, talor2019, ribas2023}. It also allows to use data acquired over several thermal cycles without the need of fitting an offset; in addition, such offsets are extremely small in the case of SPIRou. More information about the SPIRou RV zero-point is given in appendix.

The stacked periodograms shown in Figures \ref{fig:1} and \ref{fig:2} (bottom plots) illustrate how the planetary signal builds up with time, while the stellar activity signal usually seems insignificant, or fluctuating, near the expected rotation periods. This absence of RV jitter for these two M stars may be specific to nIR RV data of such slowly rotating stars, and accentuated by the use of the wPCA post-treatment.


\section{Results}
\subsection{GJ~3378}
The post-treatment removes nine data points of low S/N and thus keeps 172 measurements.
The SPIRou RV time series of GJ~3378 shows a peak at 24.7 days (Fig. \ref{fig:1}, top). Before the \wapiti\ correction, other peaks in the Bayesian periodogram \citep{mortier2015, delisle2018} are also visible, mainly near 180 and 365 days. These are typical of telluric residuals in SPIRou data and efficiently damped by the \wapiti\ algorithm (see Fig. \ref{fig:1} middle). The 24.7 day peak is untouched by this treatment and exhibits a logarithm of the Bayes factor (log $BF$) of 16.7 (a factor larger than 5 being an acceptable detection threshold). Using the $\ell_1$ periodogram \citep{hara2017} on the DACE platform \citep{delisle2016}, the most prominent peak has a Laplace Bayes factor of 65 and a period of 24.6d. After fitting a Keplerian at this period, no other peak is seen in the residuals (Fig. \ref{fig:1} bottom), in particular none near the rotation period range of GJ~3378 (80-100d) in this RV time series.

\begin{figure*}
    \centering
\includegraphics[width=0.9\hsize]{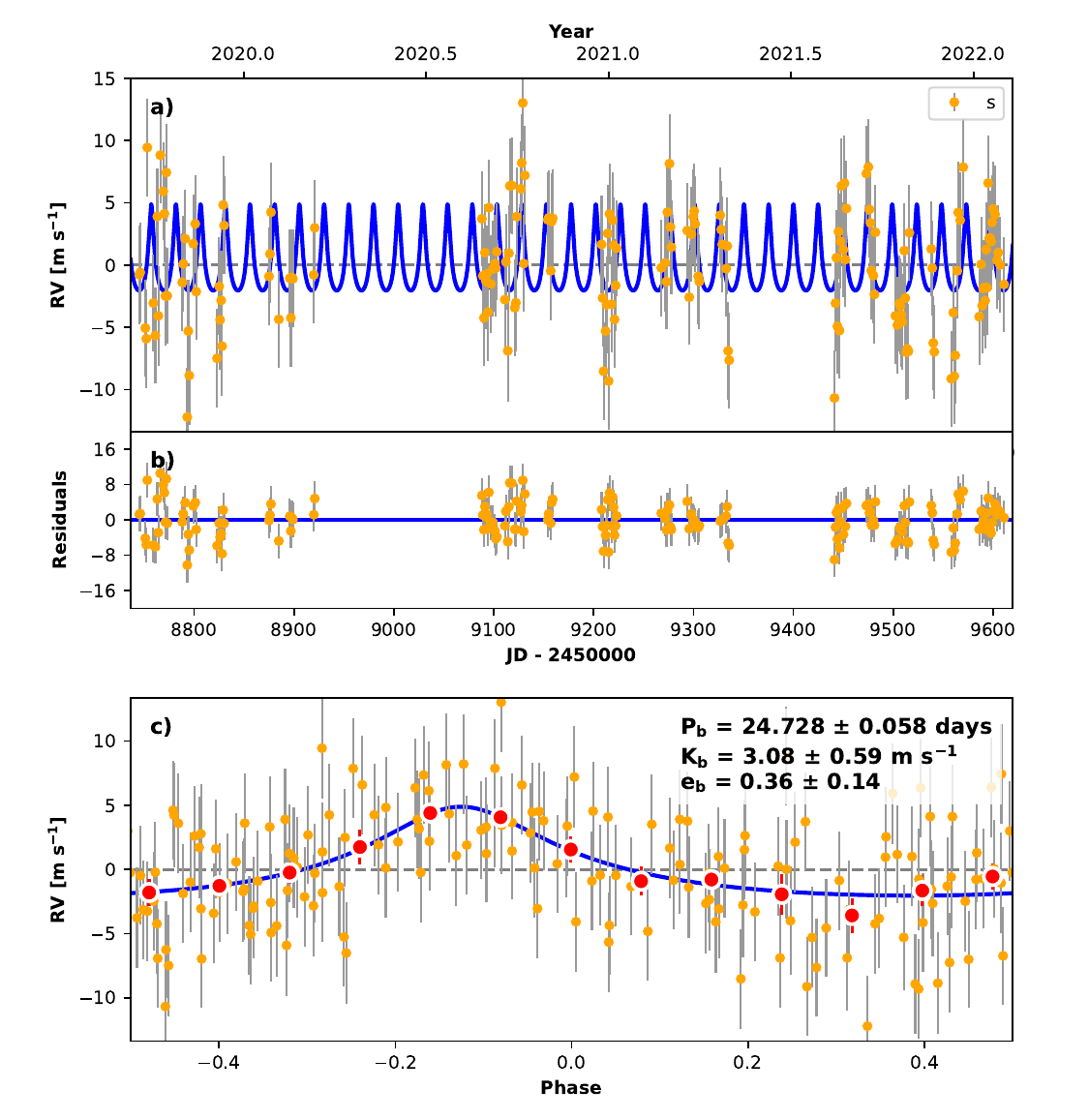}
\caption{The SPIRou RV time series of GJ~3378 as a function of time (top) and folded at the 24.7d period (bottom). The superimposed model is calculated with \radvel.}\label{fig:radvel3378}
\end{figure*}

We then used \radvel\ \citep{fulton2018} to perform the MCMC analysis of the detected signal. The procedure uses 50 walkers, 200000 burn steps and 500000 steps. 
We used the following priors: a Gaussian prior for the orbital period ($\mu=$24.7d, $\sigma=5$d), a Gaussian prior for the conjunction time ($\mu=$2459181, $\sigma=15$d) , and uniform priors for eccentricity, semi-amplitude, and white-noise jitter with bounds, respectively, to 1, 100 m/s, and 10m/s.
The preferred model encompasses an RV signal with a semi-amplitude of $3.08^{+0.6}_{-0.58}$ m/s that we interpret as due to a planetary companion of minimum mass $5.26^{+0.94}_{-0.97}$ M$_\oplus$ with an orbital period of $24.728^{+0.054}_{-0.058}$ days (Figure \ref{fig:radvel3378}). The significance of the planet signal is supported by the relative Aikike factor\footnote{https://radvel.readthedocs.io/} $\Delta$AICc of 1676. There is a preference for an eccentric orbit with eccentricity $0.36^{+0.13}_{-0.16}$. Table \ref{tab:comp} shows the model comparison numbers and Table \ref{tab:planets} lists all fitted and derived planetary parameters. A final RMS of 3.90 m/s is found while the original RMS is 4.46 m/s. Assuming a zero albedo and an atmosphere, the equilibrium temperature of the planet would be about 260~K. All parameters are listed in Table \ref{tab:planets}.

\begin{figure*}
    \centering
\includegraphics[width=0.49\hsize]{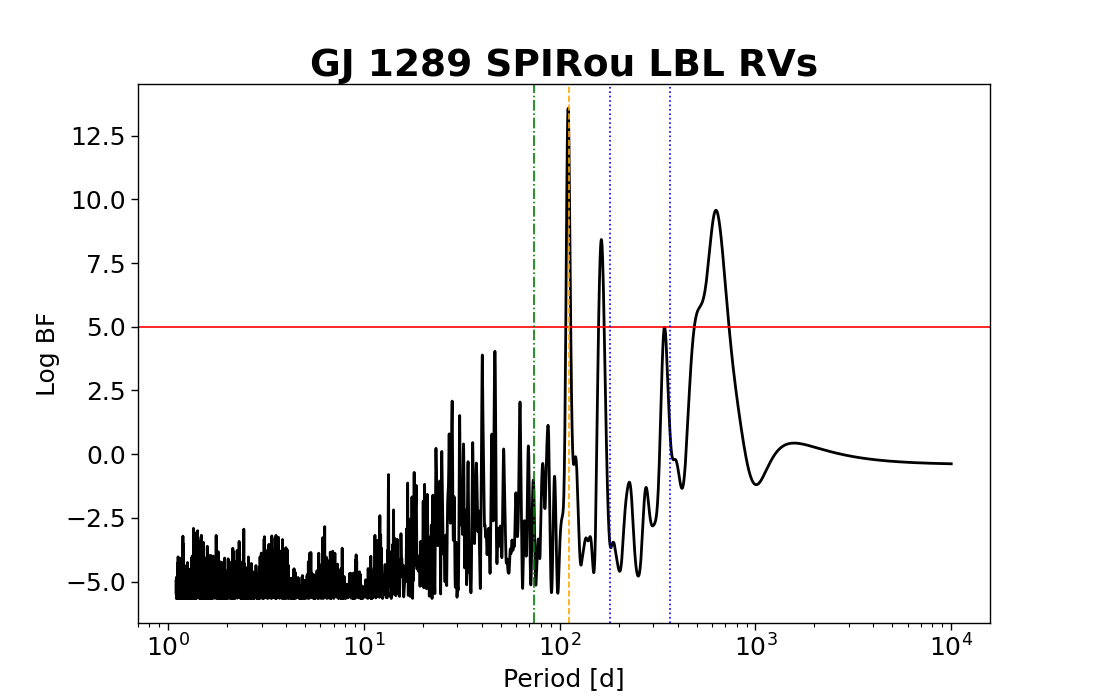}
\includegraphics[width=0.49\hsize]{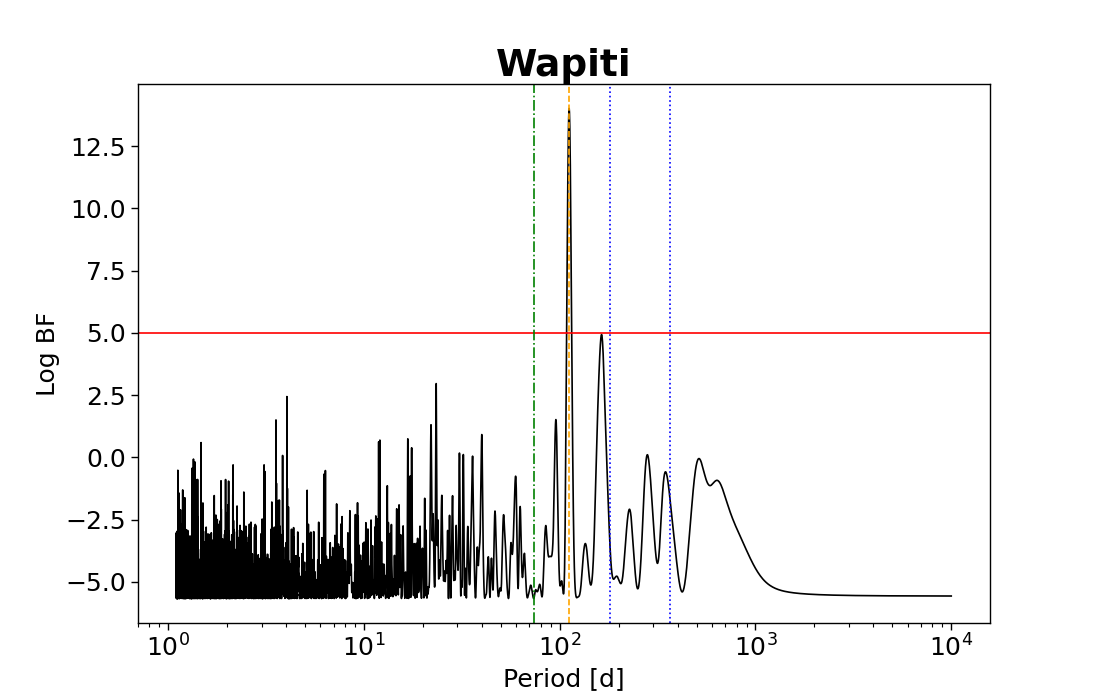}
\includegraphics[width=0.49\hsize]{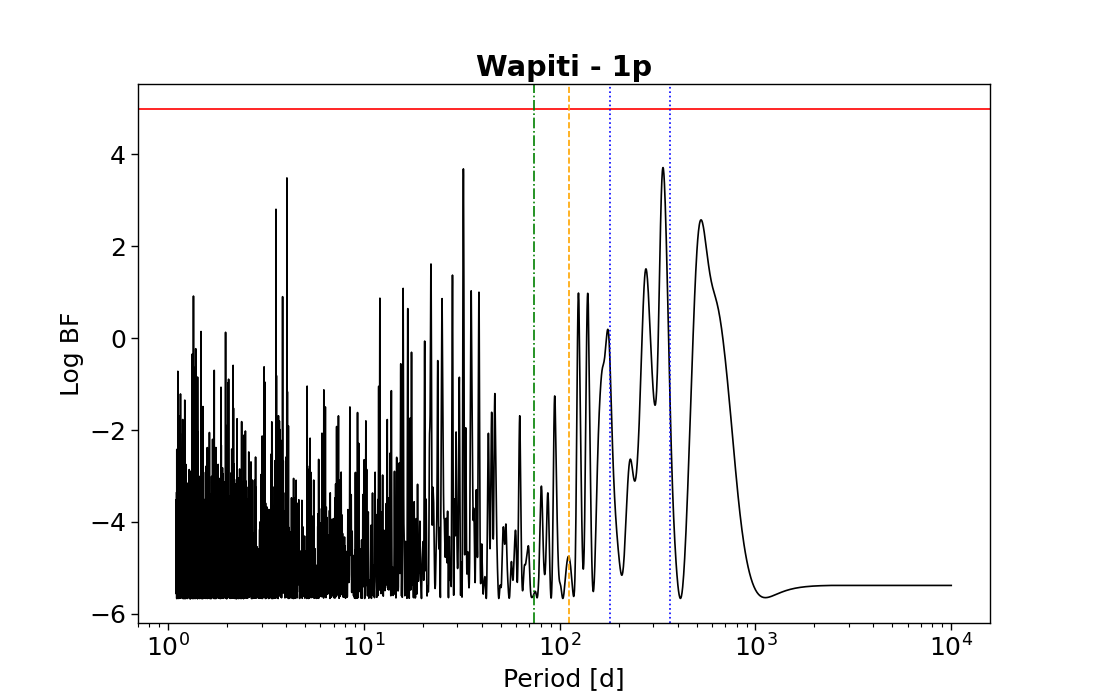}
\includegraphics[width=0.49\hsize]{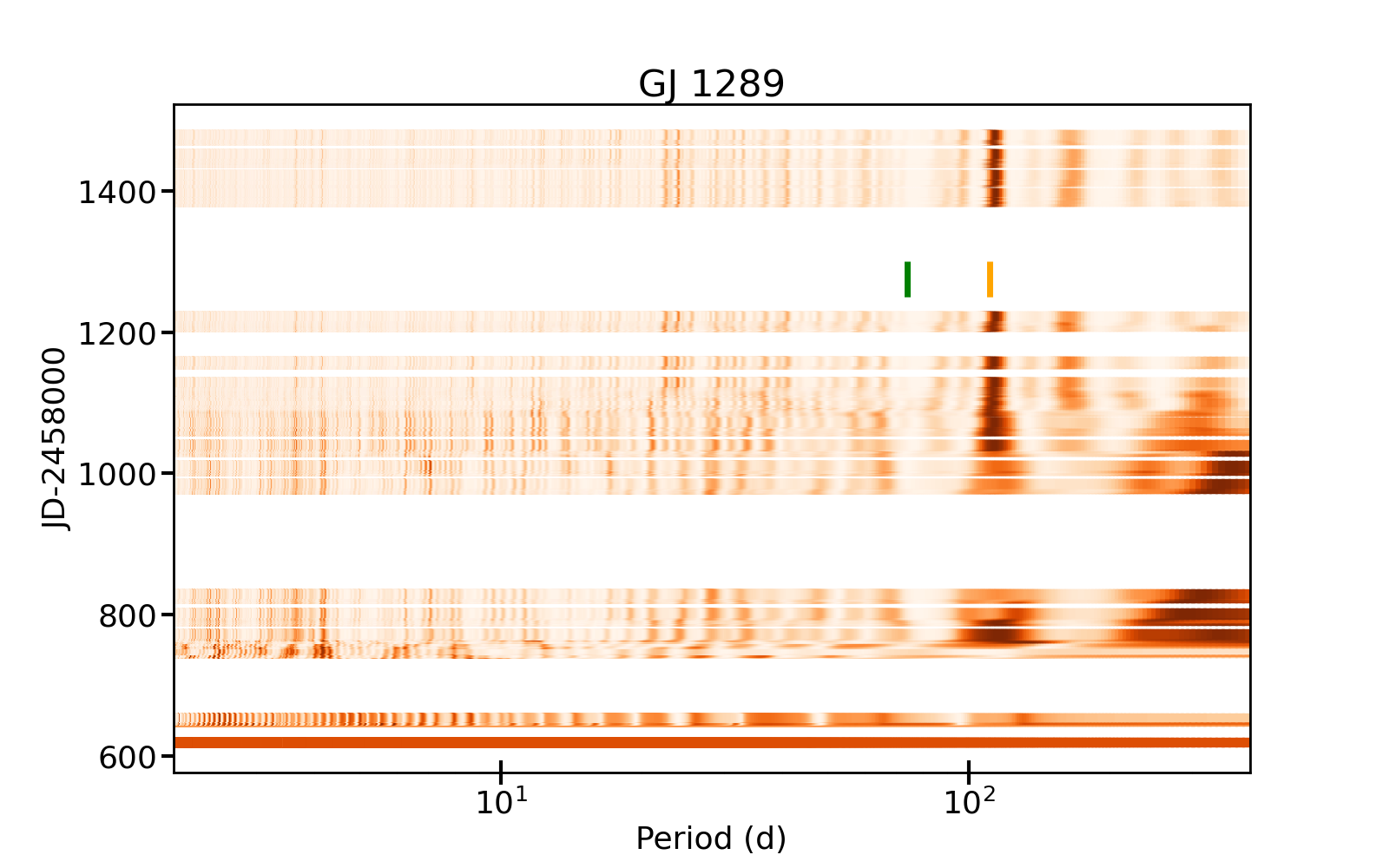}
\caption{Same as Figure \ref{fig:1} for GJ~1289.}\label{fig:2}
\end{figure*}

\begin{figure*}
    \centering
\includegraphics[width=0.9\hsize]{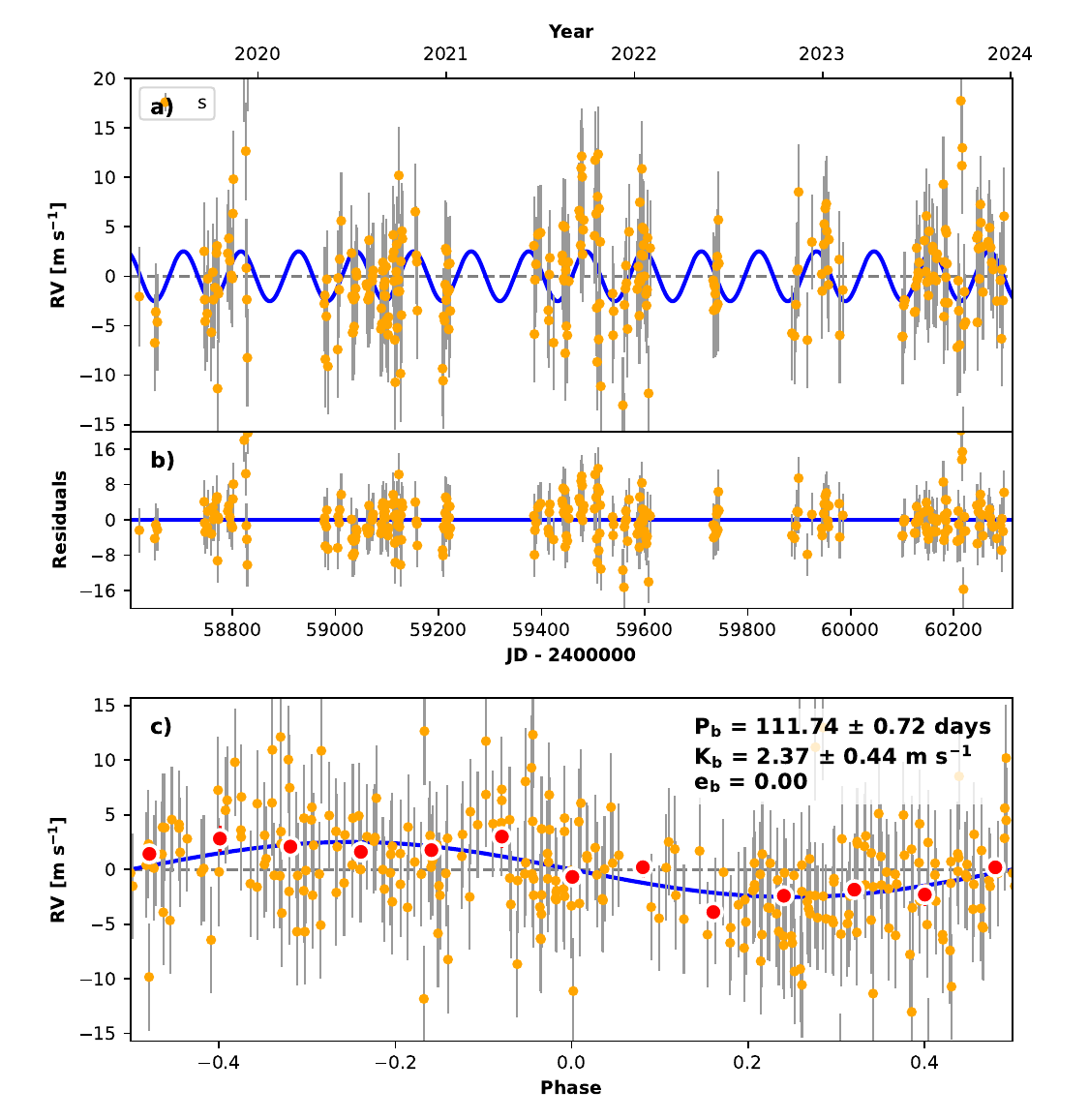}
\caption{Same as Figure \ref{fig:radvel3378} for GJ~1289.}\label{fig:radvel1289}
\end{figure*}

\subsection{GJ~1289}
The post-treatment removes eight data points of low S/N and the last 3 visits and thus keeps 289 measurements.

As for GJ~3378, we used wide priors for the MCMC modelling: a Gaussian prior for the orbital period ($\mu=$111d, $\sigma=5$d), a Gaussian prior for the conjunction time ($\mu=$2459407, $\sigma=15$d), and uniform priors for eccentricity, semi-amplitude, and white-noise jitter with bounds, respectively, to 1, 100 m/s, and 10m/s.
Using chains of 540000 steps, \radvel\ finds a more favourable one-planet Keplerian model with a companion at an orbital period of $111.74^{+0.73}_{-0.71}$ days and semi-amplitude of $2.37^{+0.43}_{-0.45}$ m/s. The signal then corresponds to a planet of minimum mass $6.27^{+1.23}_{-1.25}$ M$_\oplus$ with a preferred circular orbit. The RMS of the pre-filtered time series is 5.23 m/s while it decreases to 4.89 m/s after \wapiti\ and the Keplerian fit. A white noise jitter value of $4.69^{+0.23}_{-0.21}$ m/s results from the fit. The model comparison values are listed in Table \ref{tab:comp} and the derived parameters in Table \ref{tab:planets}.

\begin{table}
\caption{Model Comparison for GJ~3378 (top lines) and GJ~1289 (bottom lines).}
\label{tab:comp}
\begin{tabular}{lrrrr}
\hline
 & $N_{\rm free}$               & $N_{\rm data}$ & RMS & $\Delta$AICc\\\hline
 GJ 3378  $b$, eccentric                & 7 & 172 & 3.90 &  0. \\
 GJ 3378 $b$, circular                 & 5 & 172 & 4.00 &  5.30 \\
 GJ 3378, no planet                     & 2 & 172 & 4.46 &  1676. \\
\hline
 GJ 1289 $b$, circular                 & 5 & 289 & 4.89 & 0. \\
 GJ 1289 $b$, eccentric                & 7 & 289 & 4.88 & 3.52 \\
 GJ 1289, no planet                     & 2 & 289 & 5.23 & 2697. \\
  \hline
\end{tabular}
\end{table}

\section{Discussion}

CARMENES data are available for both stars in \citep{ribas2023}: 79 visits spanning 1400 days for GJ~3378 and 19 visits spanning 1600 days for GJ~1289. When using optical RVs corrected from nightly common zero-point variations of these publicly available time series, none of the reported signals are found. For GJ~1289, this is clearly explained by the small number of visits and poor sampling. There is a weak signal at 114 days in the Lomb-Scargle periodogram, although below the detection threshold. The RMS of the time series is 2.6 m/s and CARMENES data points are compatible with the signal found with SPIRou (adding CARMENES data does not change the SPIRou result). For GJ~3378, the number of visits allows a more in-depth combined analysis. The periodogram of CARMENES RVs shows low-significance peaks at 21.6, 43.9, 56.9, and 82.6 days. The first two peaks are the main ones. When fitting together SPIRou and CARMENES data, we find that: 1) the period and amplitude of the 24.7d signal are modified to $24.988^{+0.043}_{-0.044}$, i.e., a 74.5$\sigma$ difference on the period and a 3$\sigma$ difference on the semi-amplitude (1.2 instead of 3.08 m/s); 2) when fitted separately, the CARMENES RVs better fit with a pair of eccentric planets at 21.6 and 43.4 day orbital periods with semi-amplitudes of, respectively, $1.3^{+0.54}_{-0.26}$ and $1.75^{+0.68}_{-0.6}$ m/s. There is thus not a good match between CARMENES optical RVs and SPIRou nIR RVs at this stage. It is not always relevant to combine non-simultaneous optical and nIR RVs, especially at these low levels of variations, due to the chromatic variability of stars, as seen in studying more active stars like AD Leo \citep{carmona2023}. In further studies, we could combine CARMENES nIR RVs with SPIRou data, or try more complex modelling by adding other free parameters as in a Quasi-Periodic Gaussian Process modelling of optical and nIR RVs separately. This is beyond the scope of this paper which simply reports the significant RV signal seen by SPIRou, corresponding to a minimum mass of $5.26^{+0.94}_{-0.97}$ M$_\oplus$ in an eccentric orbit with an orbital period of $24.73$ days.

\begin{figure}
    \centering
\includegraphics[width=\hsize]{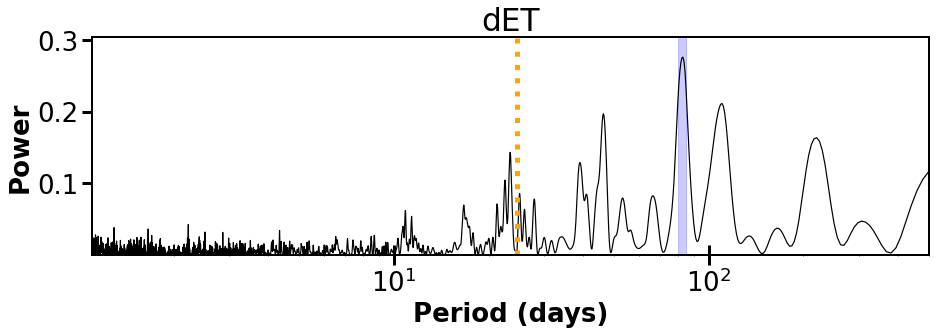}
\includegraphics[width=\hsize]{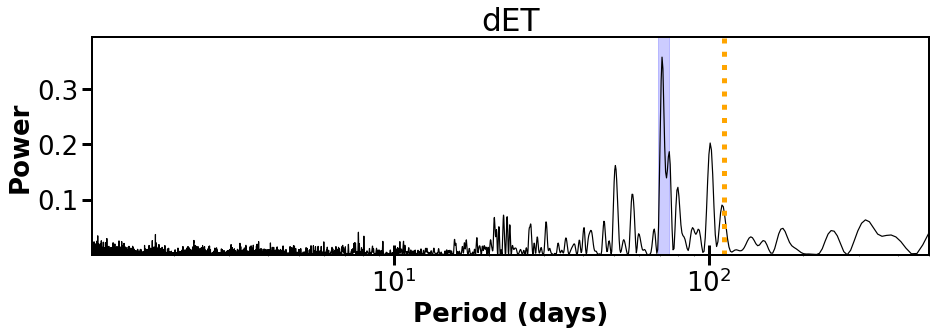}
\caption{Periodograms of the differential temperature variations for GJ~3378 (top) and GJ~1289 (bottom). Blue regions indicate the main detected periods (centered at 83d for GJ 3378 and 72d for GJ1289) while orange dashed lines show the orbital periods.}
\label{fig:dtemp}
\end{figure}

While the RV time series of both stars do not exhibit a peak in the periodogram of their RVs at the rotation period (Fig. \ref{fig:1} and \ref{fig:2}), the longitudinal magnetic field of our target stars shows a clear signature of rotational modulation with periods of 95.1$\pm$2.3 for GJ~3378 and 73.66$\pm$0.92 days for GJ~1289 \citep{fouque23, donati2023}. We also checked the behaviour of another proxy of activity, \textit{dET} (differential effective temperature), obtained with the LBL analysis. Relative temperature differences translate into tiny modulations in the thousands of lines in nIR spectra, even when RV variations are unseen; the \textit{dET} proxy appears to be one of the most robust activity indicator based on the line shape in the nIR domain (Artigau et al, in prep). The periodogram of those \textit{dET} variations for GJ~3378 and GJ~1289 is shown in Figure \ref{fig:dtemp}. The main peak for GJ~1289 is found at 71.1 days, close to the value exhibited by the magnetic field (2$\sigma$ difference). For GJ~3378, in turn, the main peak is seen at 82.9 days (a 5$\sigma$ difference from the magnetic period). This seems to confirm that the (low) signal in CARMENES may be dominated by the stellar activity at the main period of $\sim$83 days with harmonics ($\sim$43 and 21d being close to P/2 and P/4). A difference of rotation periods when using various proxies (flux/temperature contrast versus residual, large-scale magnetic field) may come from the latitudinal distribution and physical properties of active regions and the presence of differential rotation, expected to be at this level in fully-convective stars \citep{morin2010}. We attempted to model the \textit{dET} modulation observed for GJ~3378 with a quasi-periodic Gaussian Process. The RV time series was also included in the multi-dimensional fit, in order to see whether any modulation in the activity indicator would produce RV jitter. We used the \texttt{Pyaneti} package \citep{pyaneti} and use for RVs the GP and its derivative, as prescribed in \citet{rajpaul2015}. We found the following: i) the \textit{dET} modulation can be fit with the 84d period; ii) there is a significant amplitude in the GP of \textit{dET} (0.21 K) with a rather unconstrained decay time of 20-200 days and a smoothing parameter of 0.4; iii) the amplitude of the RV modulation due to the combined GP is compatible with 0 (0.3$^{+0.99}_{-1.43}$ m/s), in agreement with the fact that no peak at the rotation period or its harmonics is seen in the RV periodogram. Both GP models are shown in Figure \ref{fig:GP3378} and detailed numbers are given in the appendix table \ref{tab:GPplanets}. In summary, although the activity indicator is slightly modulated by rotation, it does not produce an RV jitter at our level of precision, for GJ~3378. 

We carried the same analysis for GJ~1289 and found again that activity does not impact the characterisation of the planet signal with SPIRou data. The amplitude of \textit{dET} is 0.97$^{+0.39}_{-0.26}$ K, while the amplitude of RV GP is 0.26$\pm0.76$ m/s.  All parameters of the exoplanet orbit agree within 1$\sigma$ error bars. This absence of RV jitter is more surprising for a star as strongly magnetic as GJ~1289. Since we are using \wapiti-corrected RVs, there is a possibility that the wPCA correction removes part of the long-term modulation that may be due to stellar activity. With rotation periods of 2-3 months, a smooth and low-amplitude modulation indeed mimics the low-frequency systematic effects that \wapiti\ is taking care of. That may explain the low amplitude of stellar activity (Ould-Elhkim et al, in prep.), although in other cases with much shorter rotation periods like AD Leo, the nIR RVs were also showing no variations \citep{carmona2023}. By looking at the whole sample of SLS data of M dwarfs, we will be able to determine in which configurations the nIR RVs shows activity modulation (as it does in the SPIRou time series of Gl 205, \citep{cortes2023}). 

We checked the $TESS$ photometric data for both stars. For GJ~3378, $TESS$ data were found in sectors 19, 59, 60, and 73. For GJ~1289, data in sectors 17 and 57 were used. For both stars, we do not report any transit detection at the expected period and phase.

\section{Conclusion}

In this paper, we report the detection and characterisation of a new planetary system orbiting the RV-quiet fully convective star  GJ~1289 of 0.21\MS, and a tentative detection of a similar planet candidate around the 0.26 \MS\ star GJ~3378. Their full properties are given in Table \ref{tab:planets}. 
These planets/candidate are the first ones found in the sample of nearby M dwarfs monitored by SPIRou \citep{moutou2023}. Other sub-programs in the SPIRou Legacy Survey have already contributed to the discovery of TOI planets and planetary companions of young stellar objects in recent years \citep{artigau2021,martioli2022,cadieux2022,donati2023a,almenara2024}.

\begin{figure}
    \centering
\includegraphics[width=1.05\hsize]{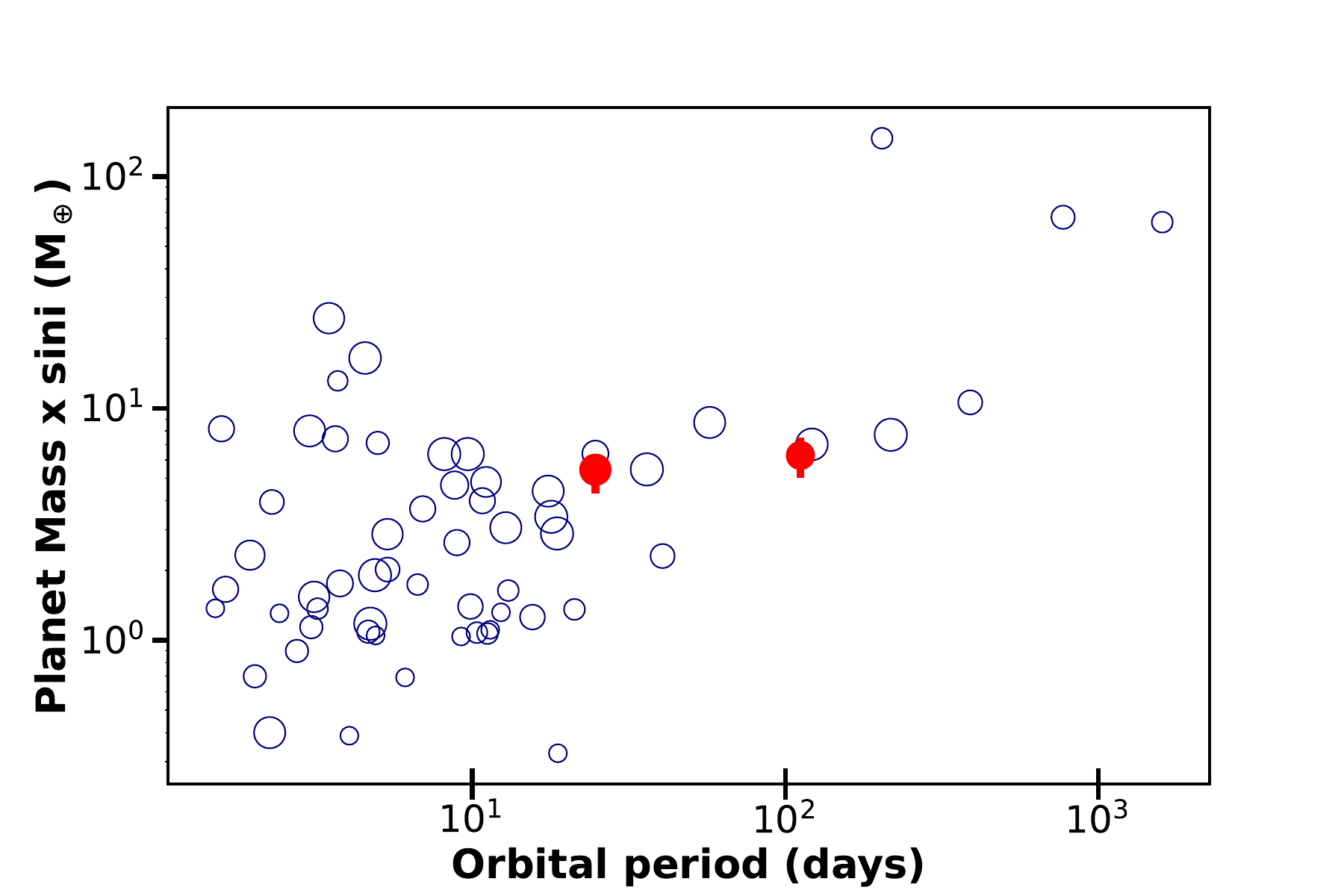}
\caption{For systems of fully-convective stars, the planet minimum mass as a function of their orbital period. The red circles represent GJ 3378 b and GJ 1289 b. The symbol size is proportional to the stellar mass, in the range from 0.09 \MS\ (Trappist-1) to 0.35 \MS\ (used limit for fully-convective stars).}\label{fig:sp}
\end{figure}

In Figure \ref{fig:sp}, the new planet candidates are compared to other known exoplanets of stars less massive than 0.35 \MS\ as function of their orbital period. There are few planets known around such stars at periods exceeding 20 days: some of such planets are found when using multiple instruments, and sometimes mixing nIR and optical RVs (like GJ 1151 c \citep{blancopozo2023}). The planet Kapteyn c (121 day orbital period and minimum mass of 7.0 \ME\ \citep{anglada2014}) lies close to GJ 1289 b in this parameter space. It is, however, a debated planet signal \citep{robertson2015, anglada2016, bortle2021} due to the possible contamination by stellar activity. As such, it is an excellent candidate for intensive nIR RV follow-up studies, as activity may be less contaminating for mid and late M stars in the nIR domain, as seen in the present study and previously compared to simultaneous optical data \citep{carmona2023}. 
The mild equilibrium temperature of sub-Neptunes in the habitable zone of their host stars make them interesting targets for future work, especially atmospheric studies; it goes with the needed caution and additional data are extremely welcome to better characterise the fundamental parameters of these new systems. It hold true especially for GJ~3378 for which non-simultaneous optical data from CARMENES do not confirm the 24.7d signal claimed here and seems rather contaminated by stellar activity. 

The conservative HZ of both host stars ranges from 0.14 to 0.28 au, while the optimistic HZ ranges from 0.11-0.30 au \citep{kopparapu2013, kopparapu2014}. This means that GJ~3378 b lies in the inner edge of the optimistic HZ while GJ~1289 b lies close to the outer edge of the conservative HZ. Both are closer to their star than the snow line, estimated to be at about 0.37 au \citep{ida2004}. As an habitable planet requires an atmosphere, being at a correct distance is not sufficient. To go further, it would be interesting to further investigate the system's evolution and architecture. The current RV data does not show evidence for giant planets further out in the system for GJ~3378, even when considering CARMENES data in addition to SPIRou data and a total time span of 6 years. Such outer giant planet could help the survival and replenishment of an atmosphere for GJ~3378 b despite the stellar evolution and activity, as shown in \citep{clement2022}. For GJ~1289 b, adding CARMENES data \citep{ribas2023} increases the time span to 7.4 years, although with scarce sampling for the first 3 years. Here again, there is no hint for a long-term companion or trend, the residual data when the 111.7d signal is removed is compatible with the jitter noise of 4.69 m/s. This level of jitter, however, seems on the high side, and it is not excluded that longer monitoring may reveal other planets in the system. In the current state, both systems do not seem similar to a resonant chain of planets as TRAPPIST-1 \citep{gillon2016,luger2017}.
The stellar masses (0.26 and 0.21\MS\, respectively for GJ~3378 and GJ~1289) are below the threshold of the sweet zone of 0.3-0.5\MS\ identified by core-accretion and migration simulations performed by \citet{burn2021}. This does not conflict the presence of planets, but may suggest that their real mass is not much larger than the measured minimum mass, as giant planets are extremely rare around low-mass stars -- although exceptions exist, as GJ 876 b \citep{delfosse1998,moutou2023} and GJ 3512 b \citep{morales2019}. 

When the Keplerian model alone is fitted to the data for GJ~3378, the best-fit eccentricity is 0.36$^{+0.13}_{-0.16}$, a 2.2$\sigma$ detection (see Table \ref{tab:planets}). When a multi-dimensional GP is used (Table \ref{tab:GPplanets}), however, the eccentricity is damped to 0.22$^{+0.15}_{-0.13}$; although compatible with the previous value within the error bar, this represents a lower significance for an eccentric orbit, at 1.5$\sigma$. We rather adopt the first determination of the orbital parameters (without GPR modeling) since stellar activity is weak and unseen in the RV periodogram (see Figure \ref{fig:1}. However, it would be useful to get more RV data and more strongly constrain the eccentricity value. If confirmed, this eccentricity could originate from several scenarii: i) gravitational interactions with a distant, more massive body in the system, ii) dynamical mean-motion resonance with a yet unseen closer planet, or iii) interactions between the planet and the disk, as described in \citep{debras2021} for more massive stars and planets. Complementary studies beyond that point would imply more RV data to better constrain other planetary companions in the system, and specific simulations of disk-planet interactions for a 0.26\MS\ star. For GJ~1289 b, the orbital eccentricity is not detected.

In addition to the interest of finding new temperate planetary systems in the solar neighbourhood, the parent stars also carry a lot of scientific motivation. Not only the level of magnetic interactions with the planets may have an impact on their habitability \citep{tilley2019}, but the lack of a tachocline in fully-convective stars may present challenges, and surprises, to the dynamo theory \citep{gastine2013, brown2020, kapyla2021}. The planet-host stars GJ~3378 and GJ~1289 have a Rossby number of, respectively, $\sim$1 and 0.7, and mean large-scale magnetic fields of about 20 and 200 G during the time of observations, with a dominating dipole. They correspond to the intermediate cases in terms of rotation, where differential rotation could be solar-like (with equators rotating faster than poles), and either cyclic or stationary, following predictions from star-in-a-box simulations \citep{kapyla2021}. A long-term monitoring of these stars with SPIRou would allow checking the cyclic behaviour, especially for the faster-evolving GJ~1289.
Observed magnetic properties may as well serve as input to estimate the magnetospheric size and shape, the stellar winds, the mass loss rate, and the magnetic energy carried over to the planet orbits. The impact on planetary atmospheric evaporation, at the relatively large orbital distance of GJ~3378 b and GJ~1289 b, is not extreme because of the slow rotation rate of the host stars \citep{vidotto2013}. Specific simulations are needed for such cases, especially for the configuration of GJ~1289 which magnetic field is evolving within a few years (by a factor of $\sim2$ until now) \citep{lehmann2024}. It could turn out that planets orbiting slow rotating low-mass stars, even when they host a strong magnetic field, are better preserved from stellar winds than more moderately active, faster rotators.

\begin{table}
\centering
\caption{Planet parameters of GJ~3378 b and GJ~1289 b}
\label{tab:planets}
\begin{tabular}{lcc}
   Parameter    &                 GJ~3378 b             &   GJ~1289 b\\\hline
  $P_{b}$ days&          $24.728^{+0.056}_{-0.06}$     &      $111.74^{+0.73}_{-0.71}$   \\
  $T\rm{conj}_{b}$ &     $2459181.2^{+1.7}_{-1.1}$        &      $2459403.2^{+3.1}_{-2.9}$              \\
  $e_{b}$ &                $0.36^{+0.13}_{-0.16}$       &       fixed (0)              \\
  $\omega_{b}$ radians &    $-0.01\pm0.51$     &      fixed (0)     \\
  $K_{b}$ (m s$^{-1}$) &       $3.08^{+0.6}_{-0.58}$       &      $2.37^{+0.43}_{-0.45}$   \\
  $\gamma$ (m s$^{-1}$)&    $0.07^{+0.31}_{-0.32}$        &      $0.07\pm 0.29$  \\
  $\sigma$ (m s$^{-1}$)&    $3.69^{+0.25}_{-0.23}$        &     $4.69^{+0.23}_{-0.21}$\\\hline
  $M_b$sini   (M$_\oplus$)   &   $5.26^{+0.94}_{-0.97}$          &    $6.27^{+1.23}_{-1.25}$  \\
  $a_b$   (au)               &   $0.106\pm0.003$        &    $0.27\pm 0.01$\\
  $T_{eq}$ (K)               &   260                    &     150 \\\hline
\end{tabular}
\end{table}

\begin{acknowledgements}
This project received funding from the European Research Council under the H2020 research \& innovation program (grant 740651 NewWorlds).
\par
E.M. acknowledges funding from FAPEMIG under project number APQ-02493-22 and research productivity grant number 309829/2022-4 awarded by the CNPq, Brazil.
\par
We acknowledge funding from Agence Nationale pour la Recherche (ANR, project ANR-18-CE31-0019 SPlaSH)and from the Investissements d’Avenir program (ANR-15-IDEX-02), through the funding of the “Origin of Life” project of the Universit\'e Grenoble Alpes.

\end{acknowledgements}

\bibliographystyle{aa}
\bibliography{00_references}

\begin{appendix}

\section{Zero-point RV time series}
\begin{figure*}
    \centering
    \vspace{-2cm}
\includegraphics[width=0.9\hsize]{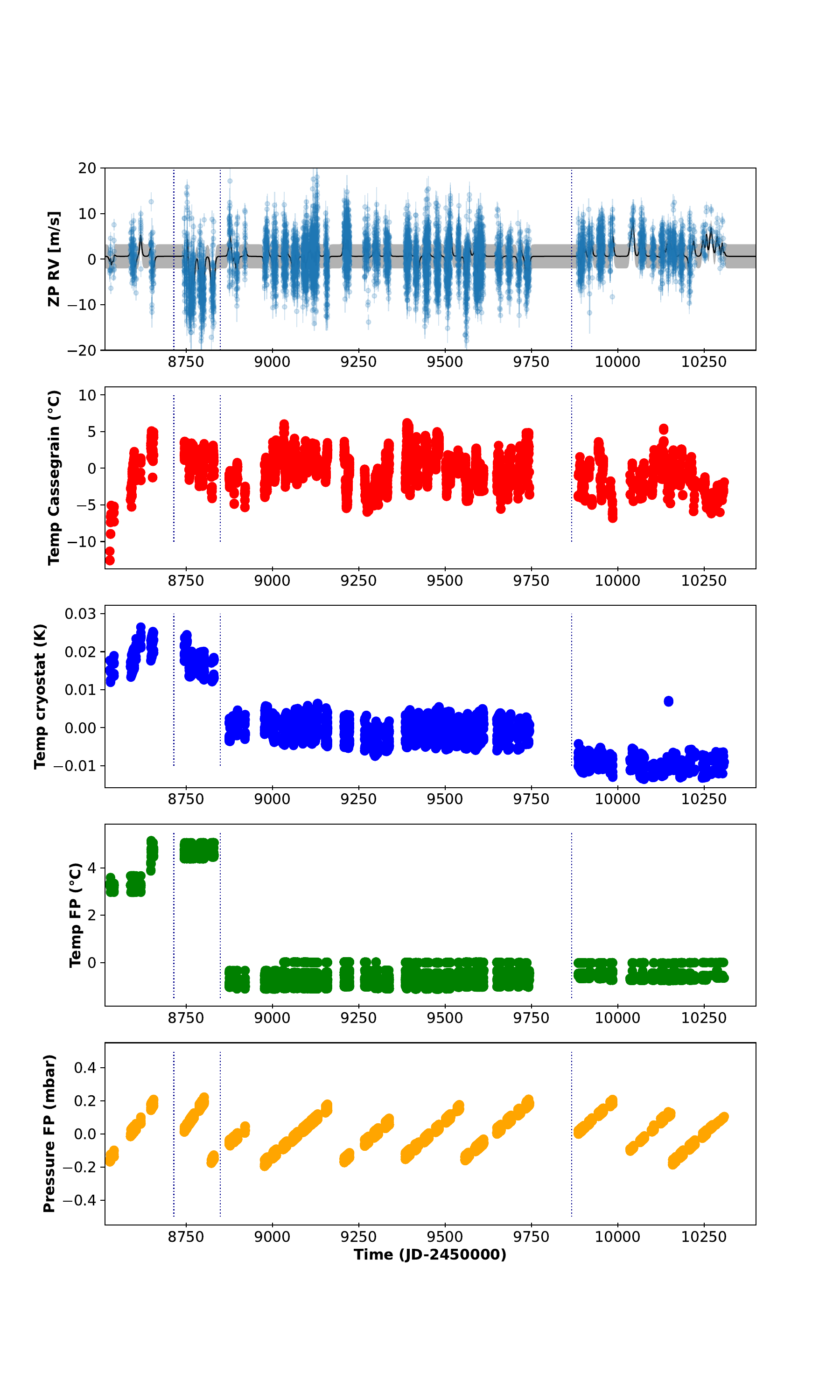}
    \vspace{-2cm}
\caption{The zero-point RV variations of SPIRou Legacy Survey data obtained with 19 RV-quiet stars (top) as function of time, as well as the evolution of a few ancillary data, from top to bottom: temperature of injection at Cassegrain, temperature of the cryostat, temperature of the Fabry-P\'erot set point, and pressure of the Fabry-P\'erot cavity (with its pumping cycle). The vertical lines show the start point of new thermal cycles. On the top plot, a Gaussian Process modeling has been applied and is used to correct for any other epoch in a continuous way.}\label{fig:zp}
\end{figure*}

The zero-point RV time series used in this analysis is shown in Figure \ref{fig:zp} as a function of time as well as the variation of a few external parameters in the same time frame. The RV data is modelled with a Gaussian Process using a Matern-3/2 kernel including a jitter term, using the \texttt{celerite} package \citep{foreman-mackey2017}:

\begin{equation}
 \kappa (\tau) = \sigma_1^2 (1 + \sqrt 3 \tau / \rho ) e^{( - \sqrt 3 \tau / \rho )} + \sigma_2^2
\end{equation}

The used solution is obtained by minimising the likelihood and has log($\sigma_1) = 0.972$, log($\rho) = 1.049$, and log($\sigma_2) = 1.128$ and log-likelihood has changed from -31309 to -21809. 

External parameters are: the temperature of the Cassegrain unit where the light is injected (mostly reflecting dome temperature), the mid-bench control temperature of the cryostat, the temperature and pressure of the Fabry-P\'erot used for simultaneous drift measurement. None of those parameters, however, correlates with the RV variations with a Pearson coefficient greater than $\sim$30\%. Statistics are shown in Table \ref{tab:statzp}. There are no jump observed during the 1245 day span of the observations. A slightly more unstable period has shortly occurred at the beginning of the survey, before Dec 2019 (corresponding to date 8849 in the plot). The set up point of the Fabry-P\'erot temperature had to be changed a few times in the early days, before the coudé room got air conditioned. The GLS periodogram of the zero-point RV variations is only characterised by the sampling, with peaks at 1 day, 29.5 days (Moon synodic period), and 1000 days (time span). 

\begin{table}
\caption{The RV dispersion of the zero-point during the SPIRou Legacy Survey. }\label{tab:statzp}
\begin{tabular}{ccc} 
 BJD-2450000  & RV dispersion (m/s) & Time span (days)  \\\hline
8515 - 8714  & 3.75$\pm$1.82 & 199 \\
8714 - 8849  & 5.12$\pm$2.53 & 135 \\
8849 - 9760  & 4.34$\pm$2.15 & 911 \\\hline
8515 - 9760  & 4.67$\pm$2.17 & 1245 \\\hline
\end{tabular}
\end{table}

\section{Corner plots}
\begin{figure*}
    \centering
\includegraphics[width=0.9\hsize]{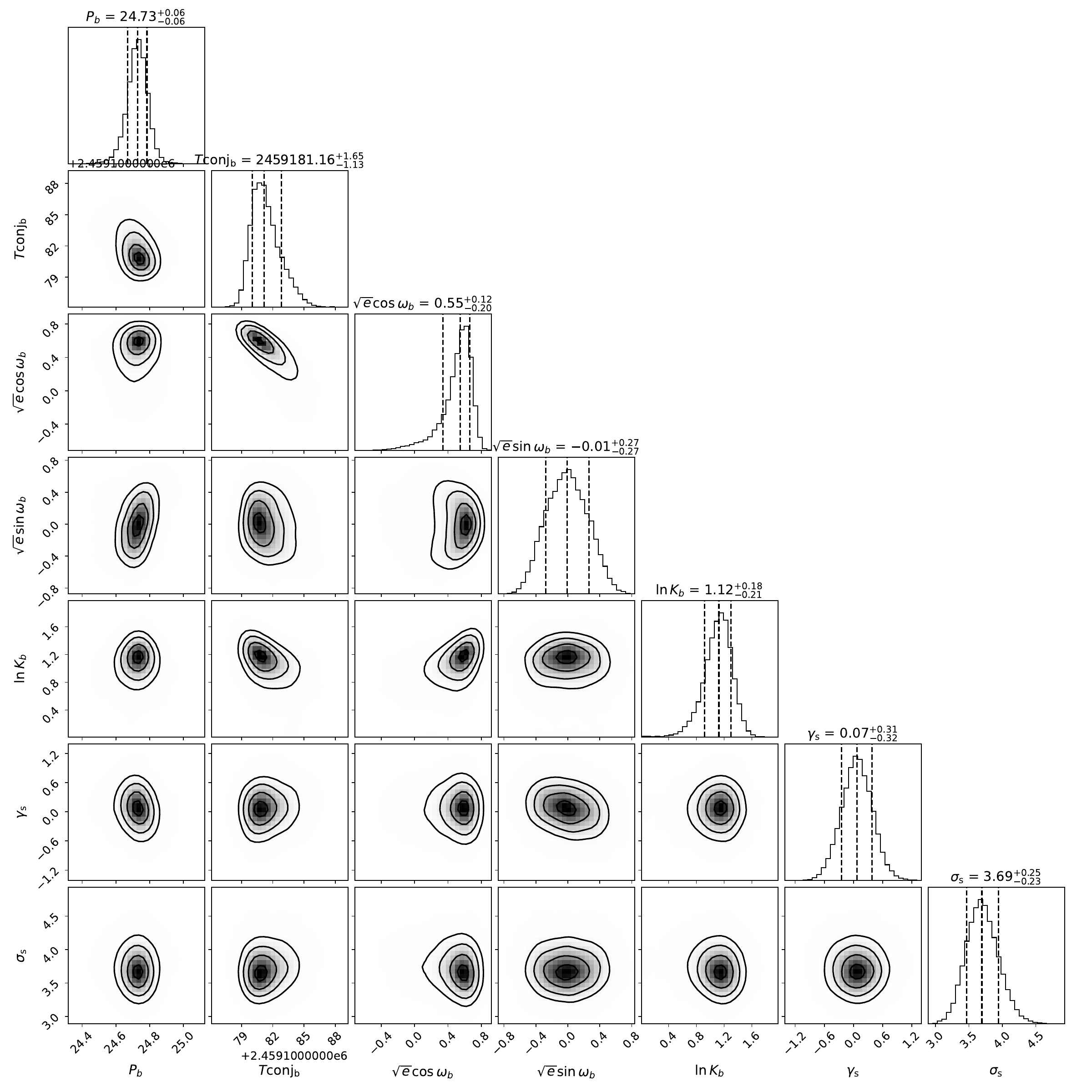}
\caption{The corner plot corresponding to the planet fit of SPIRou data for GJ~3378 (eccentric orbit) }\label{fig:corner1}
\end{figure*}

\begin{figure*}
    \centering
\includegraphics[width=0.9\hsize]{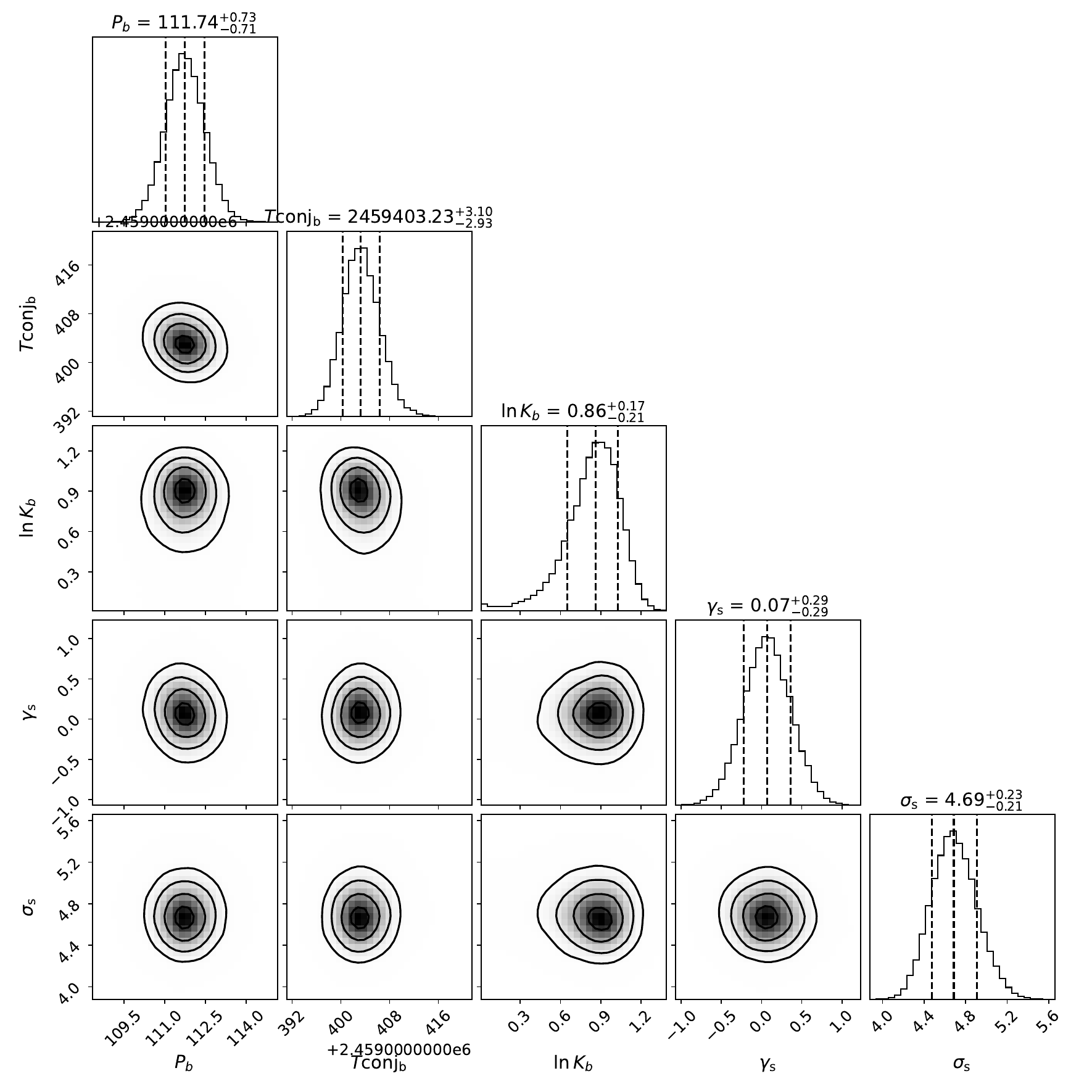}
\caption{The corner plots corresponding to the planet fit of SPIRou data for GJ~1289 (circular orbit).}\label{fig:corner2}
\end{figure*}

In Figures \ref{fig:corner1} and \ref{fig:corner2}, the corner plots of the planet fits are shown, using an eccentric model for GJ~3378 and a circular model for GJ~1289.

\section{Full solution with Gaussian Process Regression analysis}

\begin{figure*}
    \centering
\includegraphics[width=0.9\hsize]{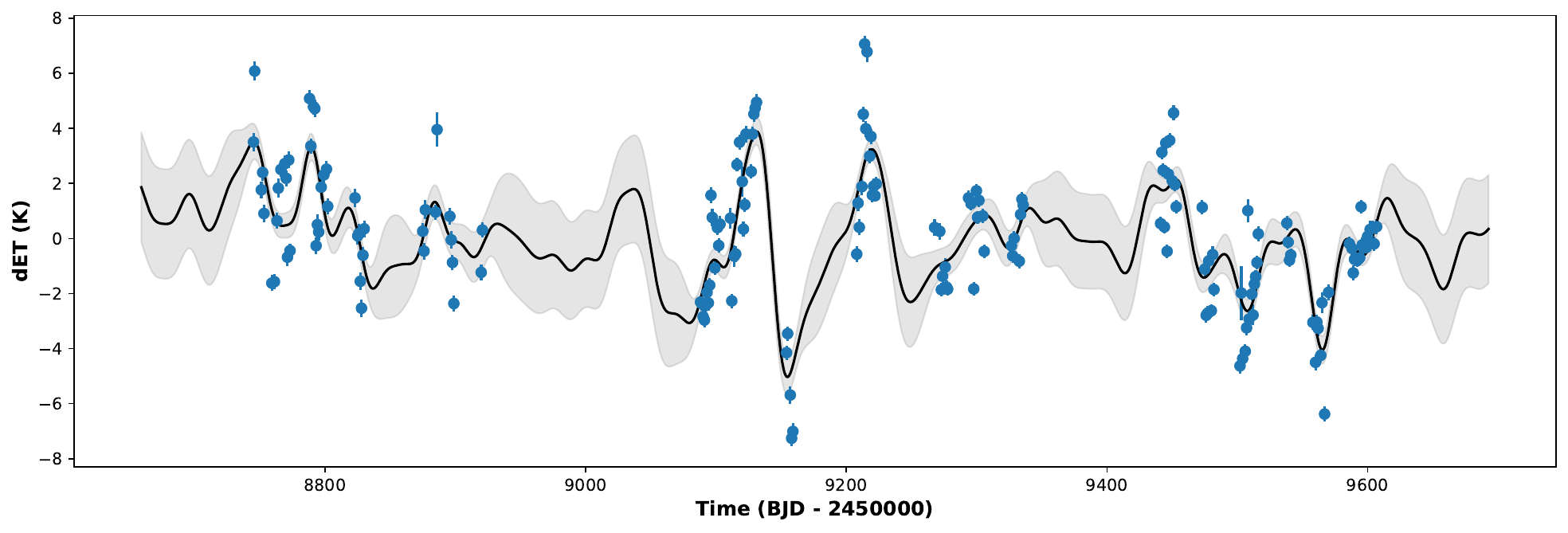}
\includegraphics[width=0.9\hsize]{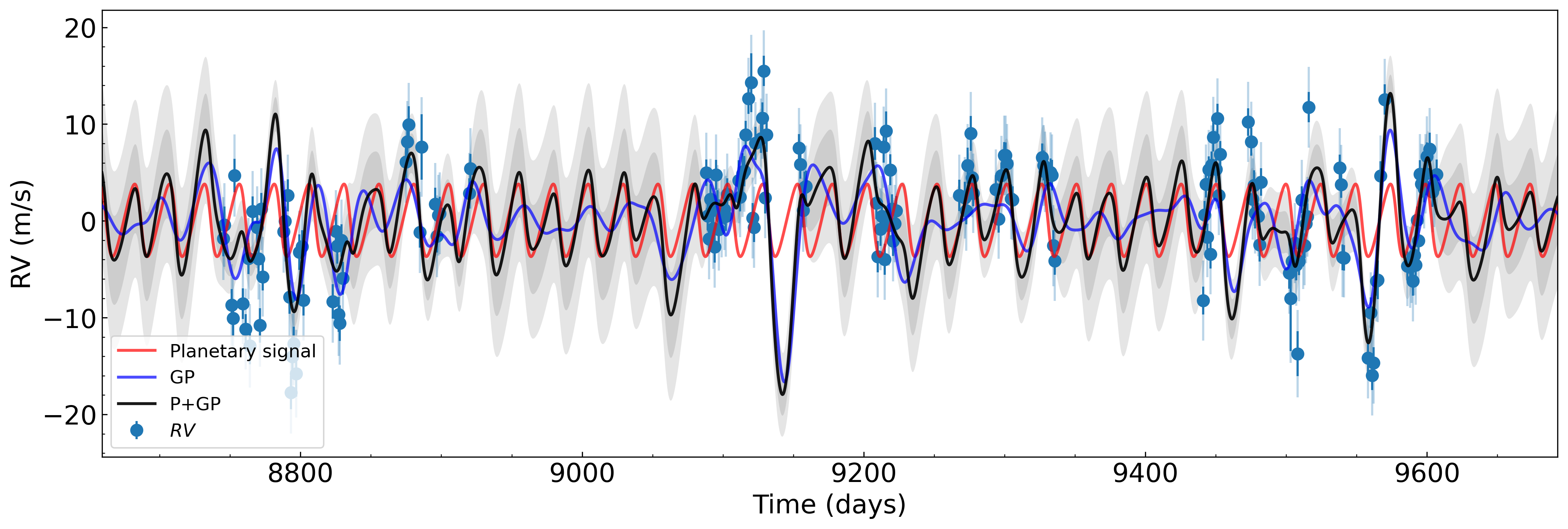}
\caption{The Quasi-Periodic GPR model (black curves) obtained for \textit{dET} (top) and for RVs (bottom) of GJ~3378, using \texttt{Pyaneti}. The Keplerian model is included in the RV modelling (red curve). SPIRou data are the blue points.}
\label{fig:GP3378}
\end{figure*}

\begin{figure*}
    \centering
\includegraphics[width=0.9\hsize]{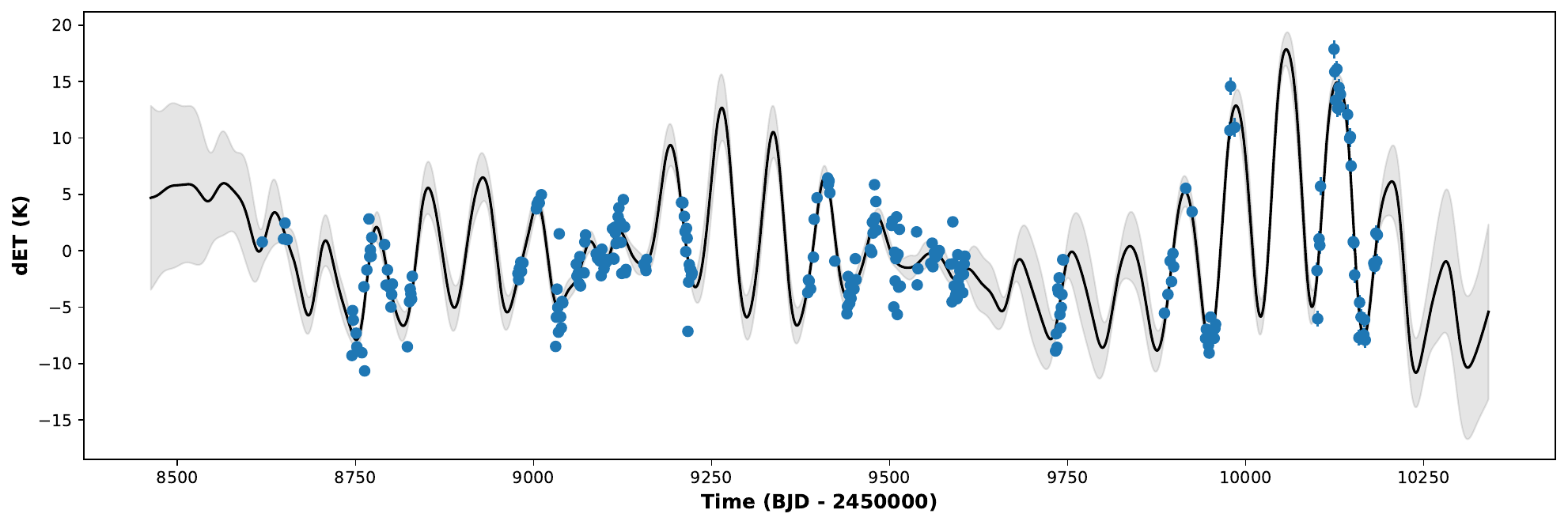}
\includegraphics[width=0.9\hsize]{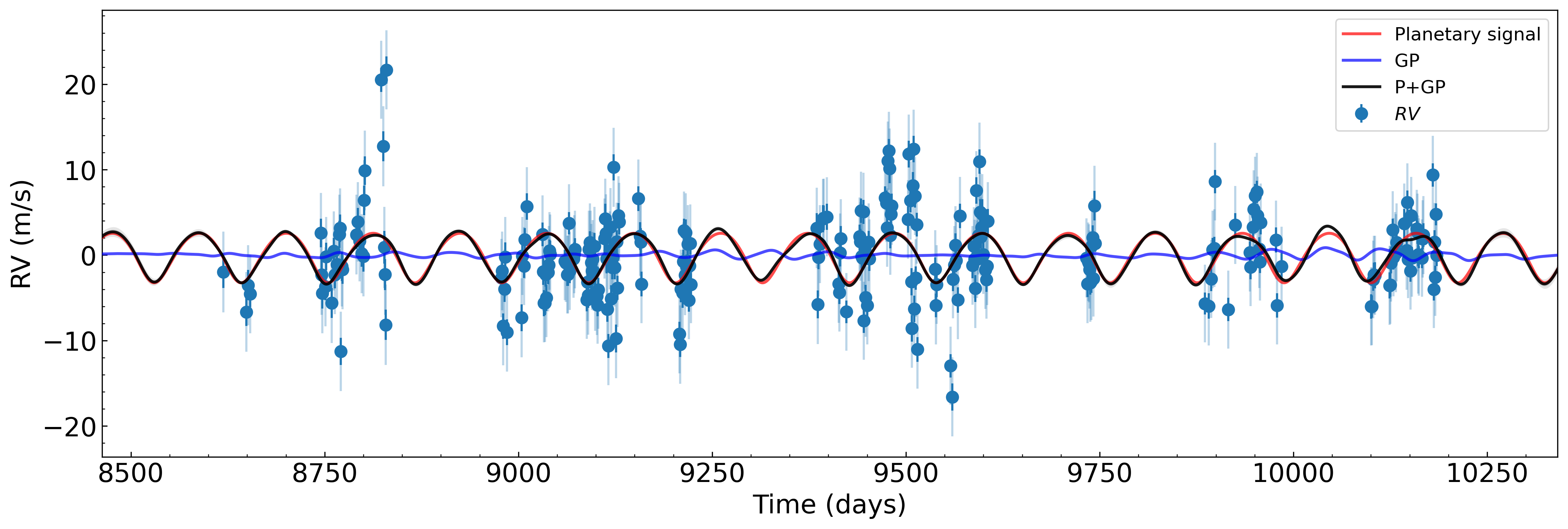}
\caption{Same as figure \ref{fig:GP3378} for GJ~1289.}
\label{fig:GP1289}
\end{figure*}

In Table \ref{tab:GPplanets}, the parameters obtained by simultaneously fitting both the Keplerian and a Quasi-Periodic Gaussian Process Regression (GPR) activity signal, using a multivariate GPR with the differential temperature \textit{dET} as an activity indicator and the RVs, are listed. The multivariate GPR model have been fitted using the \textit{pyaneti}\footnote{\hyperlink{https://github.com/oscaribv/pyaneti}{https://github.com/oscaribv/pyaneti}} model \citep{pyaneti}, briefly described by the equations \ref{eq:pyaneti} to \ref{eq:pyaneti2}.

\begin{equation}
    RV(t) = Keplerian + A_0 G(t) + A_1 G'(t) + \sigma_{RV} \label{eq:pyaneti}
\end{equation}
\begin{equation}
    dET(t) = A_2 G(t) + \sigma_{dET}
\end{equation}

\begin{equation}
    K_{i,j} = exp \left( - \frac{\sin^2 \left( {\frac{\pi \Delta t_{i,j}}{P_{rot}}}\right)}{2 \lambda_p^2} - \frac{\Delta  t_{i,j}^2}{2 \lambda_e^2}   \right) \label{eq:pyaneti2}
\end{equation}

with $RV(t)$ the radial velocities measurements; $dET(t)$ the measured differential temperature; $G$ the Gaussian Process defined with a quasi-periodic kernel with the covariance function described by equation \ref{eq:pyaneti2}, with $P_{rot}$ the stellar rotation period in days; $\lambda_e$ the evolution time scale in days and $\lambda_p$ the smoothing factor, and $G'$ is the derivative of that QP-GPR. This framework is similar to the one established by \cite{rajpaul2015}. The fitted parameters in common with the model of the Keplerian alone (Table \ref{tab:planets}) agree within errors. The models are shown in Figure \ref{fig:GP3378} and \ref{fig:GP1289}.

\begin{table}
\centering
\caption{Planet parameters of GJ~3378 b and GJ~1289 b including the joined \textit{dET} and RV GPR analysis.}
\label{tab:GPplanets}
\begin{tabular}{lcc}
\hline
   Parameter    &                 GJ~3378 b             &   GJ~1289 b\\\hline
  $P_{b}$ days&       $24.8 _{ - 3.5 } ^ { + 0.2 }$    &      $112.2 \pm 0.7$  \\
  $T\rm{conj}_{b}$ &  $2459182.1 _{ - 2.3 } ^ { + 1.5 }$ & $2459404.3 _{ - 4.3 } ^ { + 4.9 }$      \\
  $e_{b}$ &             $0.22 _{ - 0.13 } ^ { + 0.15 }$     &    $0.12 _{ - 0.09 } ^ { + 0.14 }$      \\
  $\omega_{b}$ (deg) &   $86.0 _{ - 63 } ^ { + 187 }$   &     $165 _{ - 88 } ^ { + 107 }$  \\
  $K_{b}$ (m s$^{-1}$) &  $3.8 _{ - 2.4 } ^ { + 1.3 }$     &   $2.9 \pm 0.4$   \\
  $\sigma_{RV}$ (m s$^{-1}$)&    $3.9 _{ - 0.4 } ^ { + 0.5 }$   &  $4.4 \pm 0.2$\\
  $\sigma_{dET}$ (K) &$0.15 _{ - 0.01 } ^ { + 0.02 }$ & $1.8 \pm 0.1$\\
  \hline
  GP period (d)   &   $85 \pm 5$   & $72.0 _{ - 1.8 } ^ { + 2.6 }$\\
  Decay time $\lambda_p$ (d)  & $52 _{ - 32 } ^ { + 158 }$     &$136 \pm 19$ \\
  smoothing factor $\lambda_e$& $0.43 _{ - 0.08 } ^ { + 0.55 }$    &$1.1 _{ - 0.3 } ^ { + 0.4 }$\\
  A$_0$ (km/s)   &$0.3 _{ - 1.4 } ^ { + 1.0 }$      & $0.3 _{ - 0.7 } ^ { + 0.8 }$\\
  A$_1$ (km/s d)  &$0.05 _{ - 0.1 } ^ { + 0.02 }$   & $0.006 _{ - 0.009 } ^ { + 0.01 }$\\
  A$_2$ (K)     &$0.21 _{ - 0.37 } ^ { + 0.2 }$     & $0.97 _{ - 0.26 } ^ { + 0.39 }$\\\hline
\end{tabular}
\end{table}

\end{appendix}

\end{document}